\documentclass[usenatbib]{mn2e}
\usepackage{graphicx}
\usepackage{longtable}
\usepackage{verbatim}
\usepackage{footmisc}
   \title[Observations of 23 distant JFCs]{Optical observations of 23 distant Jupiter Family Comets, including 36P/Whipple at multiple phase angles\thanks{Based on observations collected at the European Southern Observatory, Chile (proposals: ESO No. 74.C-0125, 278.C-5040), and at the Isaac Newton Group of telescopes on La Palma (proposals: I/2005A/11, I/2006A/7).} 
   }  
   
   \author[C. Snodgrass et al.]{C. Snodgrass$^{1,2}$\thanks{E-mail: csnodgra@eso.org}, S. C. Lowry$^{2}$\thanks{Now at:  NASA Jet Propulsion Laboratory, Science Division, 4800 Oak Grove Drive, Pasadena, CA 91101, USA.} and A. Fitzsimmons$^2$\\
   $^1$European Southern Observatory, Alonso de C\'ordova 3107, Vitacura, Santiago, Chile\\
   $^2$Astrophysics Research Centre, School of Physics and Astronomy, Queen's University Belfast, Belfast BT7 1NN, UK\\
      }

\begin{document}

   \date{Received <date> / Accepted <date>}

   \maketitle

   \begin{abstract}
We present photometry on 23 Jupiter Family Comets (JFCs) observed at large heliocentric distance, primarily using the 2.5m Isaac Newton Telescope (INT). Snap-shot images were taken of 17 comets, of which 5 were not detected, 3 were active and 9 were unresolved and apparently inactive. These include 103P/Hartley~2, the target of the NASA {\it Deep Impact} extended mission, {\it EPOXI}. For 6 comets we obtained time-series photometry and use this to constrain the shape and rotation period of these nuclei. The data are not of sufficient quantity or quality to measure precise rotation periods, but the time-series do allow us to measure accurate effective radii and surface colours. Of the comets observed over an extended period, 40P/V{\" a}is{\" a}l{\" a}~1, 47P/Ashbrook-Jackson and P/2004~H2~(Larsen) showed faint activity which limited the study of the nucleus. Light-curves for 94P/Russell~4 and 121P/Shoemaker-Holt~2 reveal rotation periods of around 33 and 10 hours respectively, although in both cases these are not unique solutions. 94P was observed to have a large range in magnitudes implying that it is one of the most elongated nuclei known, with an axial ratio $a/b \ge 3$. 36P/Whipple was observed at 5 different epochs, with the INT and ESO's 3.6m NTT, primarily in an attempt to confirm the preliminary short rotation period apparent in the first data set. The combined data set shows that the rotation period is actually longer than 24 hours. A measurement of the phase function of 36P's nucleus gives a relatively steep $\beta=0.060\pm0.019$. Finally, we discuss the distribution of surface colours observed in JFC nuclei, and show that it is possible to trace the evolution of colours from the Kuiper Belt Object (KBO) population to the JFC population by applying a `de-reddening' function to the KBO colour distribution.
 
\end{abstract}

 
%

\section{Introduction}

The study of short period comets has recently benefited from the successful spacecraft missions to 19P/Borrelly \citep{Soderblom02}, 81P/Wild~2 \citep{Brownlee04} and 9P/Tempel~1 \citep{Ahearn05}, which provided a wealth of information on these nuclei. The {\it Rosetta} space-craft is on its way to 67P/Churyumov-Gerasimenko \citep{Schwehm99} and will provide another great leap by studying the nucleus in detail while orbiting around it. However, these missions provide great detail on only individual nuclei, and have themselves demonstrated the great differences between the comets visited. We therefore wish to study a far larger population of cometary nuclei, in order to better understand what they have in common, and which features vary between them.

Sending space missions to a larger fraction of the cometary population is clearly impractical, and too expensive to seriously consider. Instead, we seek to measure the bulk properties of many nuclei using ground based telescopic observations. Although the activity of comets masks the nucleus when they are bright and near the Sun, it is possible to observe Jupiter Family Comets (JFCs) when they are closer to aphelion (and apparently inactive) using 2--4m class telescopes. 

A number of groups have used this method to take `snap-shot' observations of distant nuclei, allowing measurement of their brightnesses and therefore, assuming an albedo, their sizes \citep{Lowry03,Licandro00b,Meech04}. We have expanded upon this by taking time-series observations of JFC nuclei; by obtaining light-curves, constraints on the size, shape and bulk density of the nuclei can be measured. We also take multi-filter photometry in order to study the surface colours of nuclei. The first results of this work are given by \citet*{Snodgrass05} and \citet*{Snodgrass06} (hereafter papers I \& II).

In this paper we present the results of observations (or attempted observations) of 23 comets. Section \ref{ss-results} gives the results of further snap-shots obtained on comets that were not followed over an extended period, while section \ref{lc-results} gives the results of the time-series observations. We observed comet 36P/Whipple (hereafter 36P) on three separate occasions, and the results from this large data set are described in section \ref{36Psection}. Section~\ref{discussion} discusses our findings in terms of results for other comets, draws conclusions about the population of JFCs as a whole, and compares JFCs with Kuiper Belt Objects (KBOs), the supposed parent population for JFCs. Finally, section~\ref{summary} summarises all of our results.


\section{Observations and data reduction}\label{obs}

\begin{table*}
\protect\label{observations} 
\caption{Log of all observations.}
\begin{minipage}[]{\textwidth}
\renewcommand{\footnoterule}{}  

\begin{tabular}{l c c c c c c c c}        
\hline\hline
Comet & UT Date & Telescope & $r_{\rm H}$ [AU]\footnote[1]{Superscripts $I$ and $O$ refer to whether the comet is inbound (pre-perihelion) or outbound (post-perihelion). The variations in $R_\mathrm{h}$, $\Delta$ and $\alpha$ over the course of any one night were smaller than the significance quoted here.\label{fn:obs}} & $\Delta$ [AU] & $\alpha$ [deg.] & $N_{\rm exp} \times$ Filter\footnote[2]{The number of exposures $N_{\rm exp}$ in each filter in the final data set: frames rejected during reduction/analysis are not counted.\label{fn:obs3}} & $t_{\rm exp}$ [s] & Appearance\footnote[3]{Appearance: S -- Stellar, A -- Active, N -- Not detected, F -- detected but too faint to determine whether or not activity was present: these are assumed to be inactive. S$^{*}$ indicates that faint activity was later detected despite a stellar appearance.\label{fn:obs4}} \\    
\hline
8P\footnote[4]{Note that 8P/Tuttle is not a JFC, but a NIC with a short period; it has an inclination of $\sim 55\degr$ and $T_J = 1.6$.} & 01/07/05 & INT & 7.42$^I$ & 6.76 & 6.3 & 13$\times r'$ & 95 & S\\
36P & 05/03/05 & NTT & 4.08$^O$ & 3.39 & 11.0 & 5$\times$$R$, $V$, $I$ & 200 & S$^{*}$\\
& 06/03/05 & NTT & 4.08$^O$ & 3.41 & 11.2 & 17$\times$$R$, $V$, $I$ & 200 & S$^{*}$\\
& 01/03/06 & INT & 4.78$^O$ & 3.79 & 1.0 & 63$\times r'$, 10$\times V$, 10$\times i'$ & 150 & S\\
& 02/03/06 & INT & 4.78$^O$ & 3.79 & 1.0 & 55$\times r'$, 5$\times V$, 5$\times i'$ & 75 & S\\
& 12/01/07 & NTT & 5.14$^O$ & 4.94 & 11.0 & 6$\times R$, 2$\times V$ & 220 & S\\
& 26/02/07 & NTT & 5.17$^O$ & 4.35 & 6.7 & 84$\times R$, $V$ & 120 & S\\
& 27/02/07 & NTT & 5.17$^O$ & 4.34 & 6.5 & 8$\times R$ & 120 & S\\
& 18/07/07 & NTT & 5.23$^O$ & 5.45 & 10.7 & 5$\times R$, 2$\times V$ & 100 & S\\
40P & 01/07/05 & INT & 4.57$^O$ & 3.68 &  7.0 & 8$\times r'$ & 85 & S$^{*}$\\
& 02/07/05 & INT & 4.57$^O$ & 3.68 &  6.8 & 15$\times r'$, $V$, $i'$ & 85 & S$^{*}$\\
& 03/07/05 & INT & 4.58$^O$ & 3.68 &  6.6 & 7$\times r'$, $V$, $i'$ & 85 & S$^{*}$\\
& 04/07/05 & INT & 4.58$^O$ & 3.68 &  6.4 & 20$\times r'$, $V$, $i'$ & 85 & S$^{*}$\\
& 05/07/05 & INT & 4.59$^O$ & 3.67 &  6.2 & 17$\times r'$, $V$, $i'$ & 85 & S$^{*}$\\
& 06/07/05 & INT & 4.59$^O$ & 3.67 &  6.0 & 20$\times r'$, 2$\times V$, 2$\times i'$ & 85 & S$^{*}$\\
& 07/07/05 & INT & 4.60$^O$ & 3.67 &  5.7 & 18$\times r'$, 2$\times V$, 2$\times i'$ & 85 & S$^{*}$\\
44P & 01/03/06 & INT & 4.51$^I$ & 3.84 & 10.1 & 2$\times r'$ & 300 & F\\
& 02/03/06 & INT & 4.50$^I$ & 3.83 & 10.0 & 3$\times r'$, 3$\times V$, 3$\times i'$ & 170 & S\\
47P & 01/03/06 & INT & 5.11$^I$ & 4.45 & 8.9 & 2$\times r'$ & 300 & S$^{*}$\\
& 02/03/06 & INT & 5.11$^I$ & 4.43 & 8.8 & 3$\times r'$, 3$\times V$, 3$\times i'$ & 170 & S$^{*}$\\
56P & 01/03/06 & INT & 3.83$^O$ & 3.14 & 11.8 & 3$\times r'$, 2$\times V$ & 520 & A\\
& 02/03/06 & INT & 3.83$^O$ & 3.15 & 12.0 & 5$\times r'$, $V$, $i'$ & 260 & A\\
70P & 01/07/05 & INT & 4.84$^I$ & 4.26 & 10.6 & 6$\times r'$ & 295 & F\\
72P & 02/03/06 & INT & 3.28$^O$ & 2.35 & 7.1 & 5$\times r'$ & 50 & N\\ 
75P & 02/07/05 & INT & 4.95$^I$ & 3.94 & 1.7 & 3$\times r'$ & 75 & N\\
78P & 02/03/06 & INT & 3.84$^O$ & 2.86 & 1.8 & 3$\times r'$ & 60 & A\\ 
94P & 04/07/05 & INT & 4.14$^O$ & 3.19 & 5.6 & 14$\times r'$, $V$, $i'$ & 75 & S\\
& 05/07/05 & INT & 4.14$^O$ & 3.18 & 5.3 & 20$\times r'$, 3$\times V$, 2$\times i'$ & 75 & S\\
& 06/07/05 & INT & 4.14$^O$ & 3.18 & 5.1 & 26$\times r'$, 2$\times V$, 2$\times i'$ & 75 & S\\
& 07/07/05 & INT & 4.15$^O$ & 3.18 & 4.9 & 23$\times r'$, 2$\times V$, 2$\times i'$ & 75 & S\\
103P & 01/03/06 & INT & 5.03$^O$ & 4.30 & 8.3 & 2$\times r'$, $V$, $i'$ & 200 & A\\
& 02/03/06 & INT & 5.03$^O$ & 4.29 & 8.1& $r'$, $V$, $i'$ & 110 & A\\
104P & 05/03/05 & NTT & 3.06$^O$ &  2.31 & 13.8 & 2$\times$$R$ & 80 & N \\
114P & 04/07/05 & INT & 3.75$^I$ & 2.95 & 10.9 & 8$\times r'$ & 90 & S\\
120P & 02/03/06 & INT & 3.89$^O$ & 3.05 & 8.8 & 5$\times r'$ & 115 & F\\ 
121P & 01/03/06 & INT & 3.92$^O$ & 3.43 & 13.5 & 4$\times r'$, $V$, $i'$ & 400 & S\\
& 02/03/06 & INT & 3.93$^O$ & 3.42 & 13.4 & 18$\times r'$, $V$, $i'$ & 215 & S\\
& 31/05/06 & FTN & 4.18$^O$ & 3.34 & 8.6 & 14$\times R$, 3$\times V$ & 60 & S\\
131P & 02/03/06 & INT & 3.48$^O$ & 2.65 & 10.1 & 3$\times r'$, 3$\times V$ & 105 & S\\ 
135P & 02/03/06 & INT & 3.55$^I$ & 2.65 & 7.5 & 5$\times r'$ & 80 & N\\ 
160P & 02/03/06 & INT & 3.98$^O$ & 3.41 & 12.5 & 3$\times r'$ & 140 & F\\ 
P/1995 A1 & 02/07/05 & INT & 5.52$^I$ & 4.71 & 6.9 & 8$\times r'$ & 135 & S\\
2006 BZ8\footnote[5]{2006~BZ8 was observed to search for activity in this newly discovered object, which has a comet-like ($T_J = -1$) orbit. None was found.} & 02/03/06 & INT & 2.32$^I$ & 1.35 & 7.1 & 15$\times r'$, 3$\times V$, 3$\times i'$ & 10 & S\\
P/2004 H2 & 01/07/05 & INT & 3.70$^O$ & 3.06 &  13.6 & 14$\times r'$ & 210 & S$^{*}$\\*
& 02/07/05 & INT & 3.70$^O$ & 3.06 &  13.4 & 19$\times r'$, $V$, $i'$ & 210 & S$^{*}$\\
& 03/07/05 & INT & 3.71$^O$ & 3.05 &  13.3 & 14$\times r'$, $V$, $i'$ & 210 & S$^{*}$\\
& 04/07/05 & INT & 3.71$^O$ & 3.04 &  13.1 & 16$\times r'$, 2$\times V$, $i'$ & 210 & S$^{*}$\\
& 05/07/05 & INT & 3.71$^O$ & 3.03 &  12.9 & 14$\times r'$, 4$\times V$, 2$\times i'$ & 210 & S$^{*}$\\
& 06/07/05 & INT & 3.72$^O$ & 3.02 & 12.7  & 10$\times r'$, 6$\times V$, 3$\times i'$ & 210 & S$^{*}$\\
& 07/07/05 & INT & 3.72$^O$ & 3.02 & 12.6 & 11$\times r'$, 4$\times V$, 2$\times i'$ & 210 & S$^{*}$\\
P/2004 H3 & 05/07/05 & INT & 3.71$^O$ & 2.97 & 12.0 & 6$\times r'$ & 90 & F\\
P/2004 T1 & 02/03/06 & INT & 3.82$^O$ & 2.86 & 3.9 & 5$\times r'$ & 60 & N\\ 
P/2001 X2 & 02/07/05 & INT & 5.02$^I$ & 4.02 & 2.4 & 2$\times r'$ & 80 & N\\
\hline                                   
\end{tabular}
\\[-5ex]

\normalsize
\end{minipage}
\end{table*}

The data presented were taken using three different telescopes over 5 observing runs of varying lengths. A log of all the observations is given in table \ref{observations}. The majority of the data were taken with the 2.5m Isaac Newton Telescope (INT) at the Roque de Los Muchachos observatory on the island of La Palma. 

The INT was used during two runs. The first was in July 2005, on seven nights (July 1st -- 7th) around the time of the {\it Deep Impact} mission's encounter with 9P/Tempel~1. The telescope was being used to follow the evolution of the dust coma and tails of this comet before and after the impact \citep{lowry-DI,Jones05}, but was available during the later half of each night after 9P had set. Conditions were excellent and all seven nights were photometric; time-series data were collected on three comets [40P/Vaisala~1, 94P/Russell~4 and P/2004~H2 (Larsen)] and snap-shot observations were made of 7 others. The telescope was used again between the 27th of February and the 2nd of March 2006, during which the first two nights were lost due to severe winter weather. The third night was clear but non-photometric, while the 4th night was photometric with periods of excellent seeing (the $FWHM$ of the image PSF varied between 0.8 and 2.7\arcsec, with a median of 1.1\arcsec). Time series observations of comets 36P/Whipple and 121P/Shoemaker-Holt 2 were obtained, and snap-shots were taken of 12 other objects.

The imager on the INT is the Wide Field Camera (WFC) which is mounted at the primary focus. The WFC is a mosaic of four thinned EEV 2048$\times$4096 pixel CCDs, which has a pixel scale of 0.33\arcsec{} per pixel and a total field of view of  34\arcmin{} on a side. The wide field nature of the camera is not necessary for this work, and only one element of the multi-extension fits file is used to minimise the reduction workload. In the 2006 data this mostly corresponds to the central CCD~4, while the comets were centred on CCD~3 in the 2005 data as the telescope pointing had been offset to make use of the full width of the WFC mosaic in observing 9P's tails. Even the single CCD has a field of view of 11.5$\times$23 arcmin$^2$, which gives the advantage of having a constant set of comparison stars over two nights for all but the fastest moving comets, allowing light-curves to be fully differential with no uncertainties from night to night calibration. Also, by a fortuitous coincidence, comets 44P/Reinmuth 2 and 47P/Ashbrook-Jackson were close together on the sky during the February/March 2006 run, and the wide field of the WFC allowed both to be observed simultaneously giving `two for the price of one'. 44P fell on CCD~1 and 47P fell on CCD~2. The filters used were the Harris $V$ and Sloan $r'$ and $i'$. The Sloan filters have the advantage of having sharp cut-on and -offs with little overlap in sensitivity at wavelengths between the bands, although further reduction steps are required as the $i'$-band has significant fringing which had to be subtracted from the data manually for each frame, as pipelines or automated routines failed to measure the fringe brightness correctly.

We used the 3.6m New Technology Telescope (NTT) at the European Southern Observatory's La Silla site on the nights of the 5th to 7th March 2005. This telescope, the EMMI instrument used, the run and the associated data reduction are described in paper II, where the majority of results from it were published. The data taken on 36P during that run were not included in paper II as it is more sensibly grouped with the other observations of this comet described here. The NTT and EMMI were also used in the 3rd time-series on 36P, taken on the nights of the 26th and 27th of February 2007. The conditions were clear, with seeing measured in the images between 0.6 and 1.3\arcsec{}, with a median of 0.8\arcsec{} FWHM. 

The fifth set of observations was carried out using the 2.0m robotic Faulkes Telescope North (FTN). We used approximately 1 hour on the 31st of May 2006 to perform follow up observations on comet 121P/Shoemaker-Holt~2, which INT observations had shown to be brighter than expected and showing evidence of having only recently ceased out-gassing (see section \ref{121Psect}), to search for any sign of continuing activity. The comet was imaged in the Bessell $R$-band using the DillCam CCD imager, which has a single 2048$\times$2048 pixel CCD with a pixel scale of 0.278 arcsec per pixel in the default 2$\times$2 binning mode, giving a 4.6$\times$4.6 arcmin$^2$ field of view. Frames were also taken of the comet through the Bessell $V$-band, and of standard stars through both bands, to allow for photometric calibration. 

The reduction for the NTT, FTN and 2006 INT data and photometry on all frames were performed using standard IRAF tasks as described in papers I \& II. The 2005 INT data were reduced (along with the 9P data taken during the run) using the Wide Field Survey (WFS) pipeline\footnote{{\tt http://www.ast.cam.ac.uk/$\sim$wfcsur/}}; inspection of data from each run showed that both reduction methods gave equally good results. 

Differential photometry provides a precise measurement of how the brightness of the comet varies with time by comparing it to the brightness of nearby field stars, which can be assumed to have constant intrinsic brightness. Calibration of these differential magnitudes onto a standard scale was achieved by taking observations of \citet{Landolt} standard star fields, and solving for zero points, extinctions and colour terms in the usual way. These were used to calculate the magnitudes of the field stars in each frame; taking the mean of these values gave a very accurate measurement of the brightness of the comparison stars, $\overline{m_R(s)} \pm \sigma_R(s)$. The full calibrated comet magnitude is then given by
\begin{equation}
m_R(c) = \left[m'_R(c) - m'_R(s)\right] + \overline{m_R(s)} + k'_R\delta(V-R)
\end{equation}
where the term in brackets is the differential instrumental magnitude (comet - star), and $\delta(V-R)$ is the difference in $(V-R)$ colour between the star and the comet. When the colour term $k'_R$ is low (as with the NTT observations with $|k'_R| \le 0.06$ during both runs) and the comparison stars and the comet have similar colours, the final term is negligible. Care was taken to select comparison stars with colours in the range expected for cometary nuclei to minimise any uncertainties in this colour term, although experimentation found this method of obtaining calibrated photometry to be robust and independent of which comparison stars were chosen. In the INT data, with Sloan $r'$ and $i'$ filters, the colour terms are significant for even slight differences between star and comet colours. As can be seen, the dependance on the comet colour in the calibrated magnitude causes the determination of both the magnitude and the colours to be an iterative process. A direct calibration of the comet photometry was used to provide a first estimate of its colour (with large error bars), and therefore calibrated magnitudes and from those a revised $(V-R)$ colour. This method typically found a stable solution within three or four iterations.

It is important to note that while the calibration onto standard magnitude scales is necessary for comets in different star fields, and to allow calculation of absolute magnitudes and colours for the objects, searching for periodicity can be done just as well using differential light-curves. Where possible the differential light-curve is preferred as it contains no sources of error beyond the photometric measurement uncertainties, allowing optimum measurement of the relative variation of brightness with time.


\section{Results from snap-shot imaging}
\label{ss-results}

`Snap-shot' images of all potential target comets were taken. There are three possible outcomes from imaging of comets; the comet is either undetected, unresolved or active: Here we present the snap-shot data divided into three sections according to these appearance categories. Undetected comets are either simply too faint to see, in which case upper limits on the size are found, or confused with background sources, in which case very little information can be obtained. Unresolved comets are detected and appear as point sources, and may be interpreted as being bare nuclei, depending on evaluation of the contamination by dust comae. Those which appeared to be unresolved, bright and well separated from background stars were followed as time-series (see section \ref{lc-results}); results on the remainder are presented here. Active comets have resolvable coma in individual images; limits are placed on the level of activity and the size of the nucleus. All these results are summarised in table \ref{results_snapshots}. 

\begin{figure*}  
  \begin{tabular}{c c c}
   \includegraphics[width=0.3\textwidth]{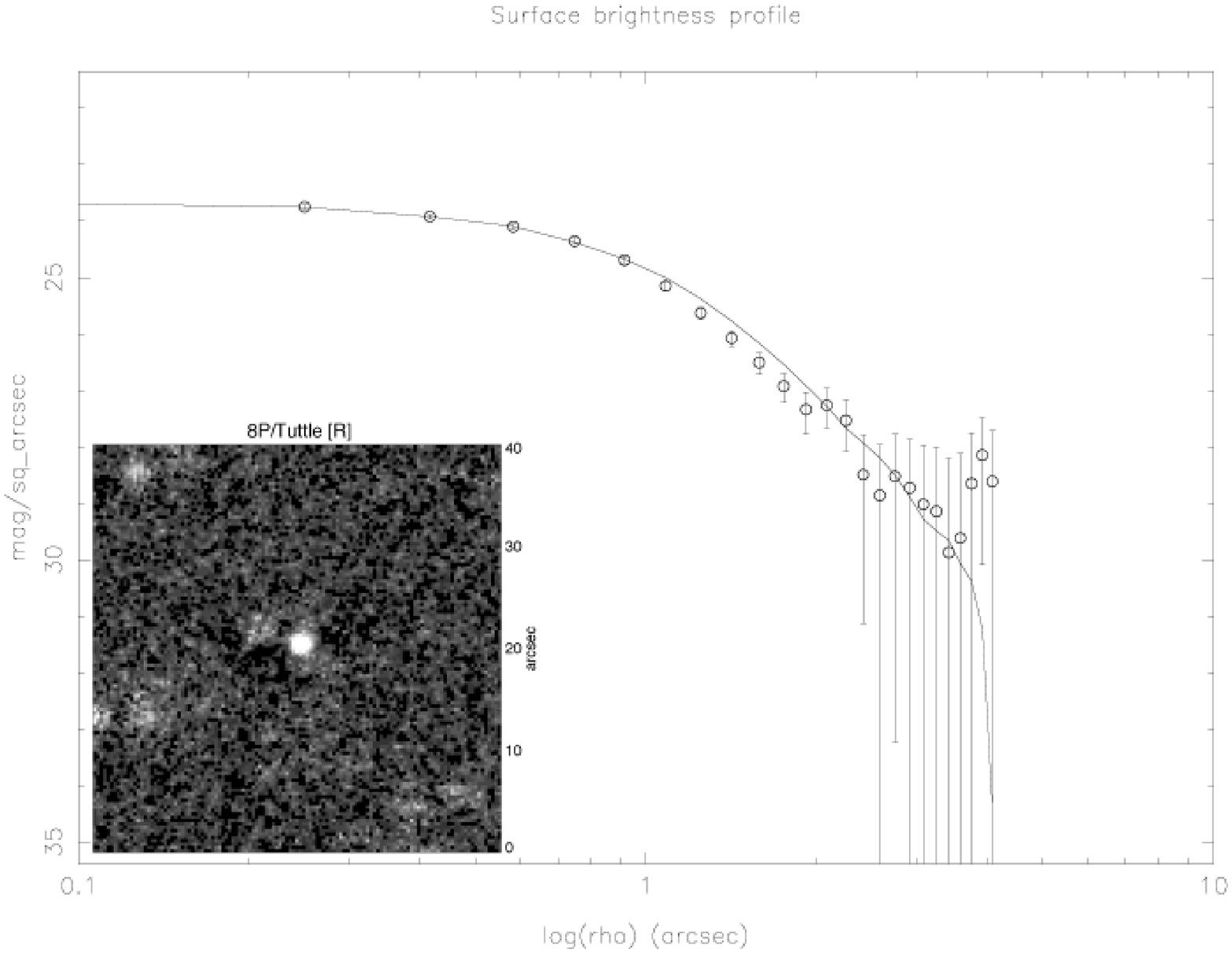} &
    \includegraphics[width=0.3\textwidth]{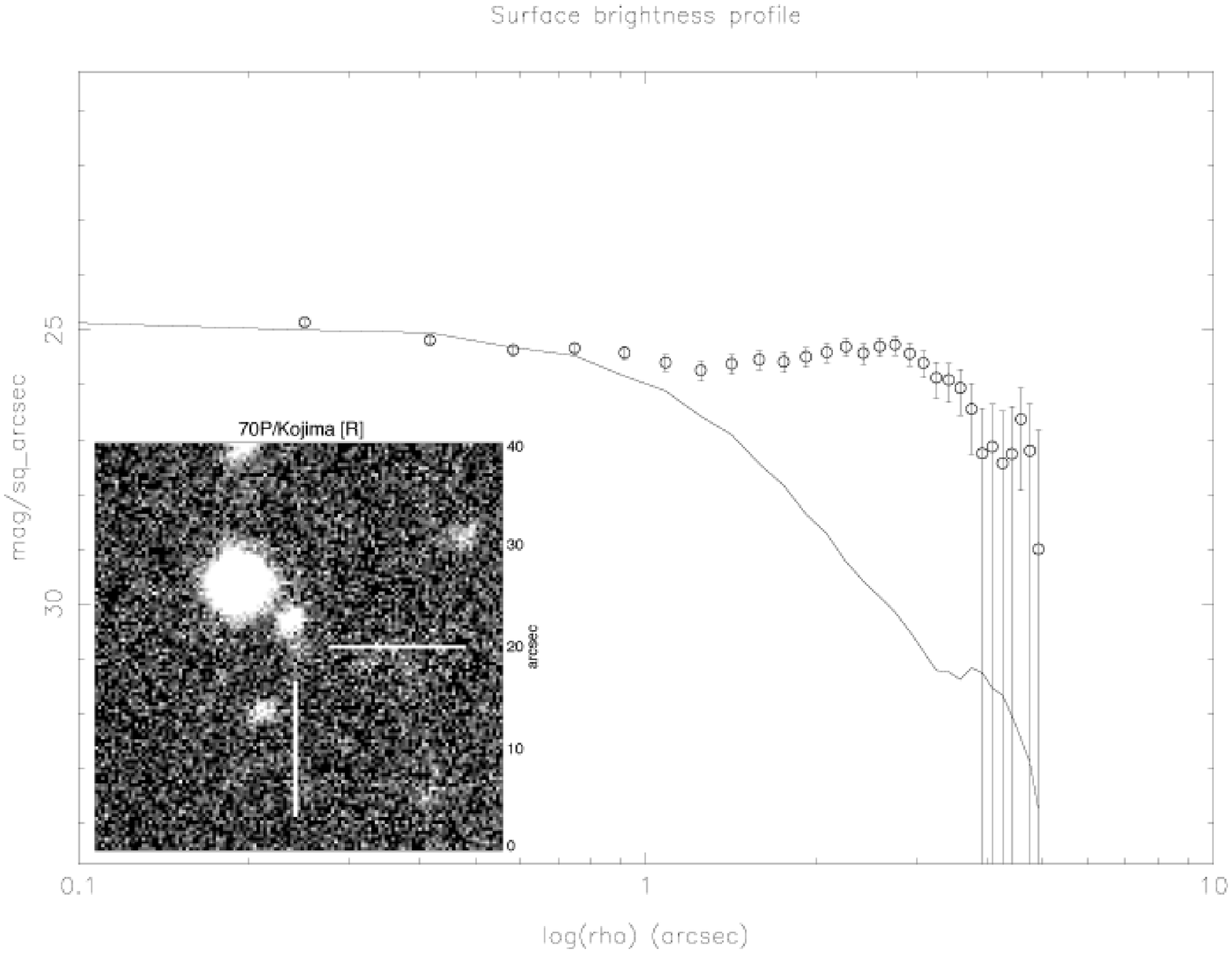} &
   \includegraphics[width=0.3\textwidth]{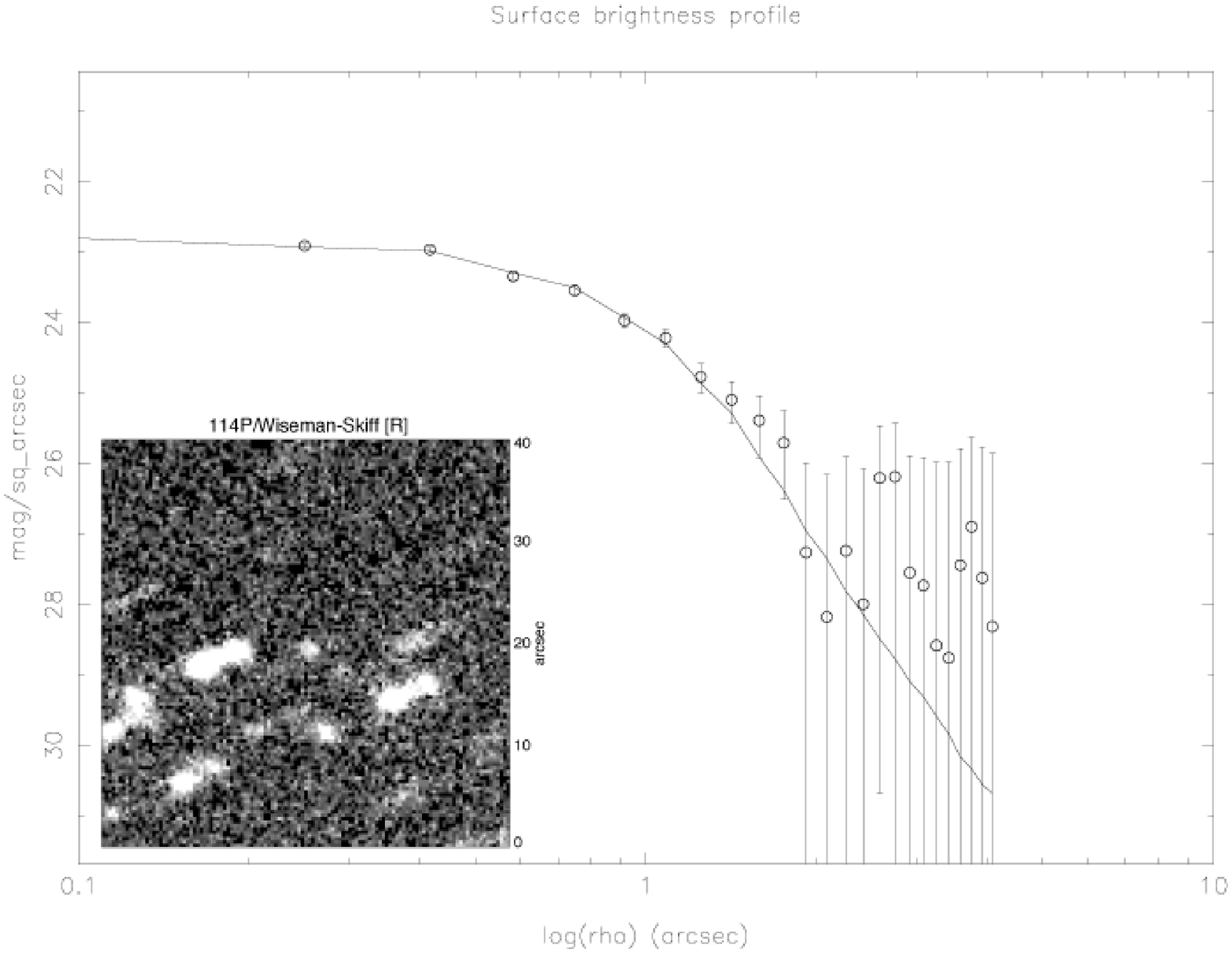} \\
   \includegraphics[width=0.3\textwidth]{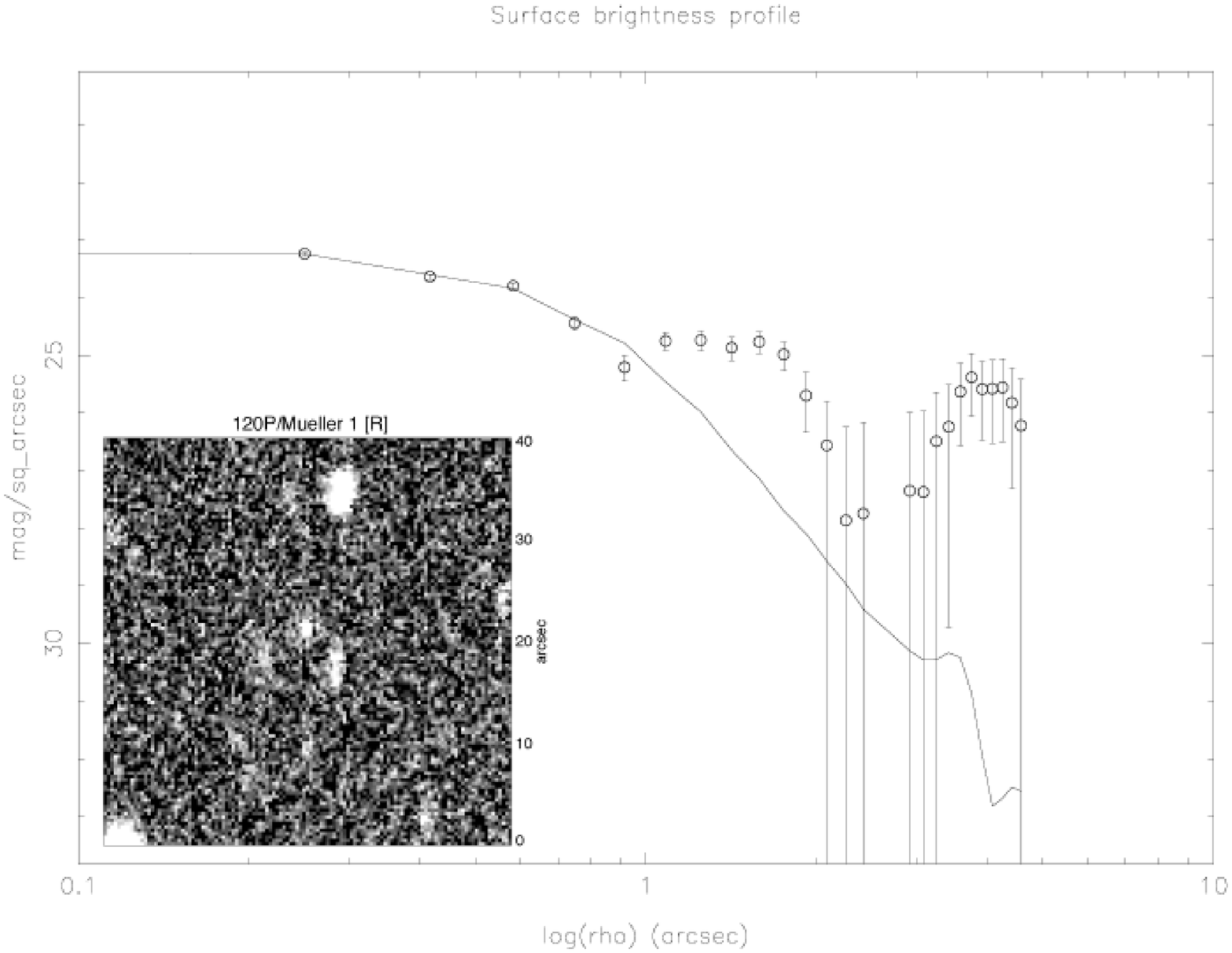}  &
   \includegraphics[width=0.3\textwidth]{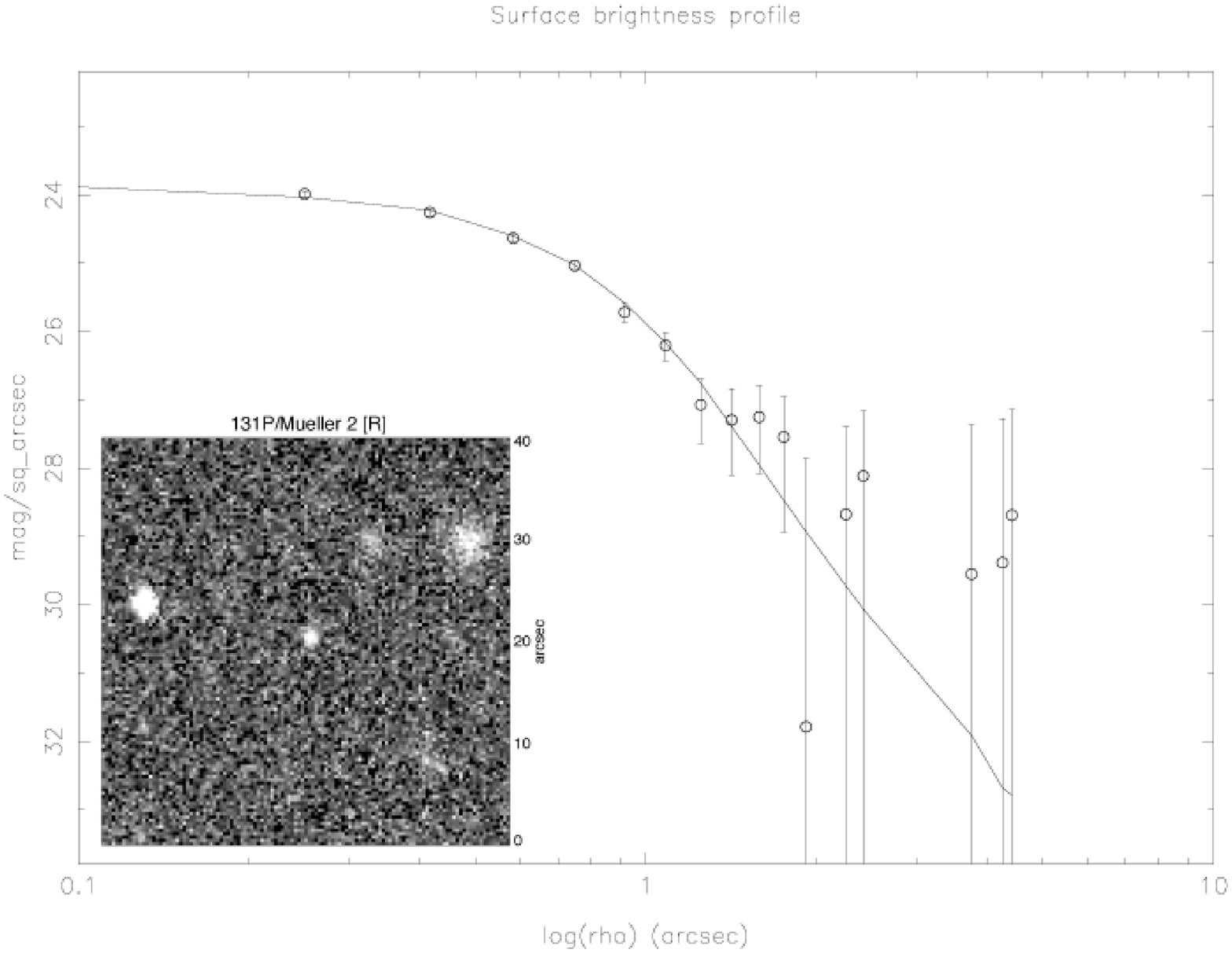} &
   \includegraphics[width=0.3\textwidth]{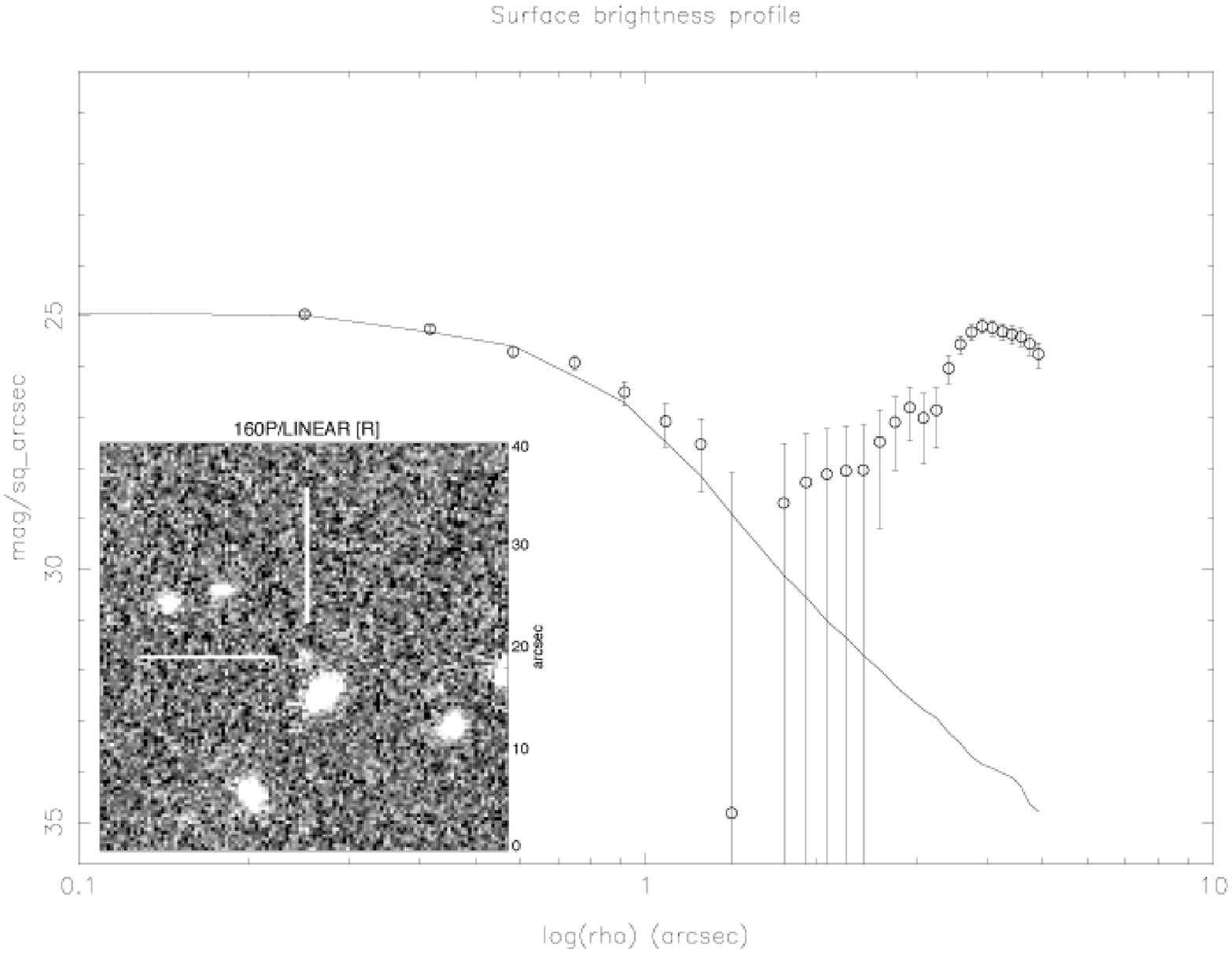} \\
   \includegraphics[width=0.3\textwidth]{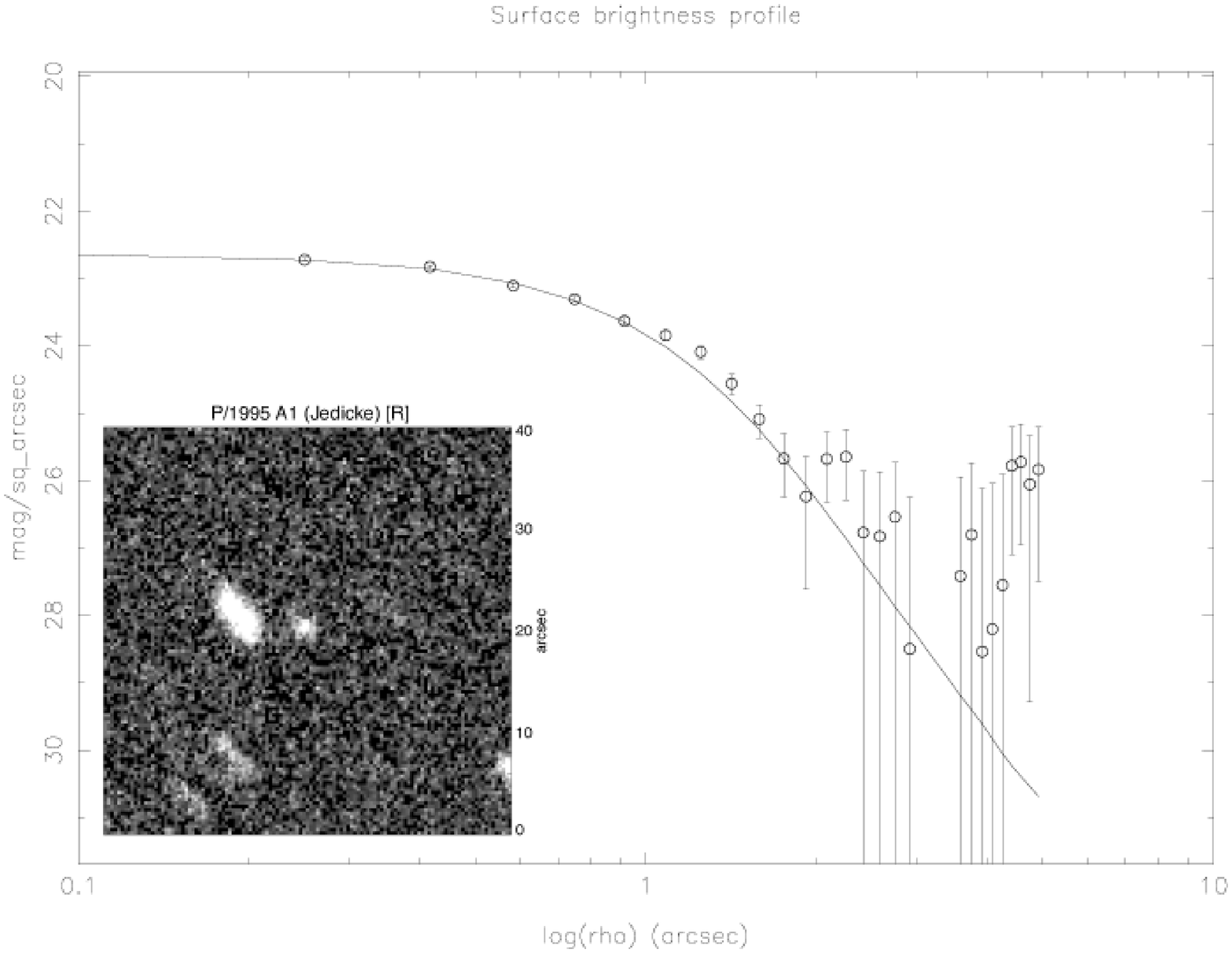} &
   \includegraphics[width=0.3\textwidth]{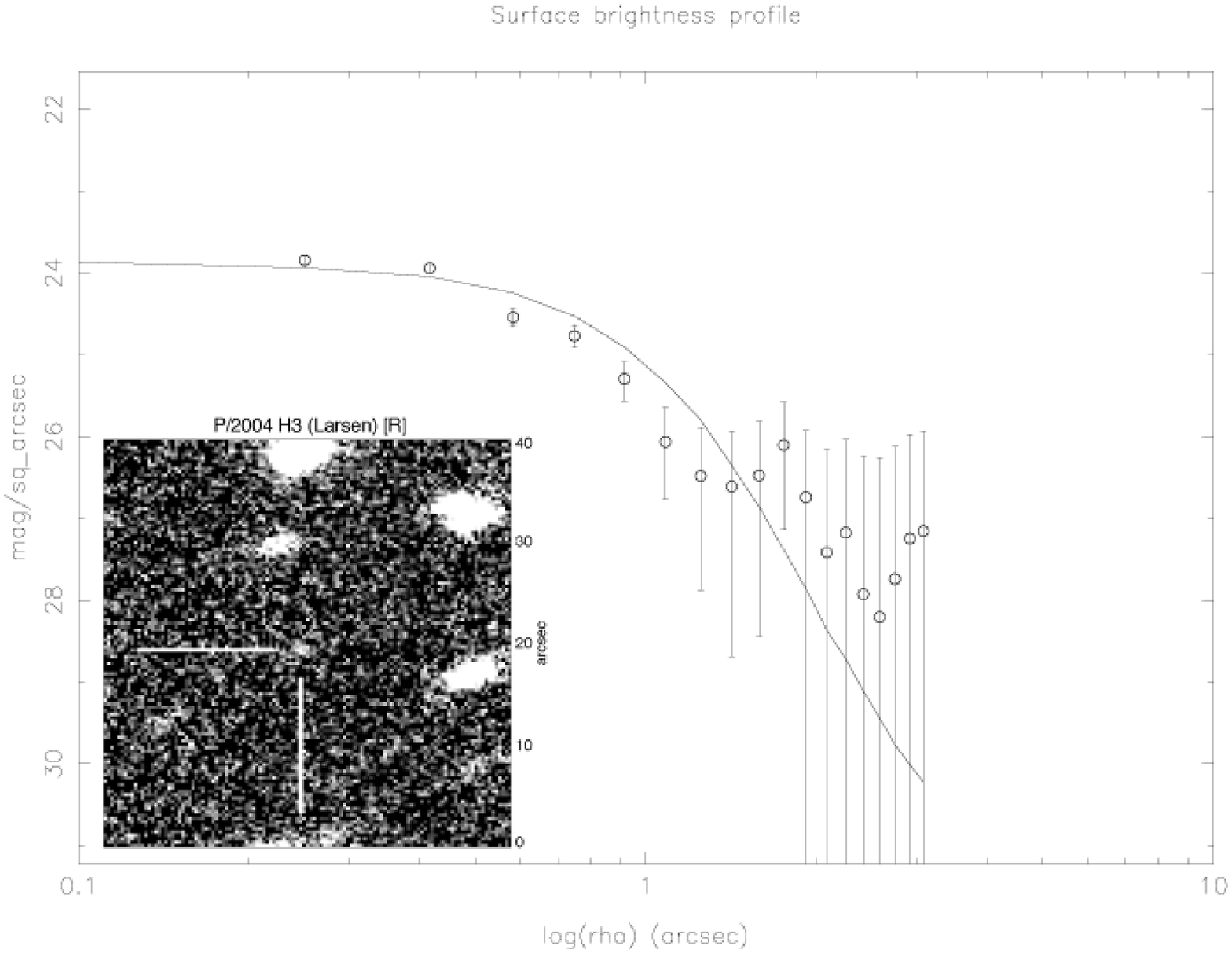} &
   \includegraphics[width=0.3\textwidth]{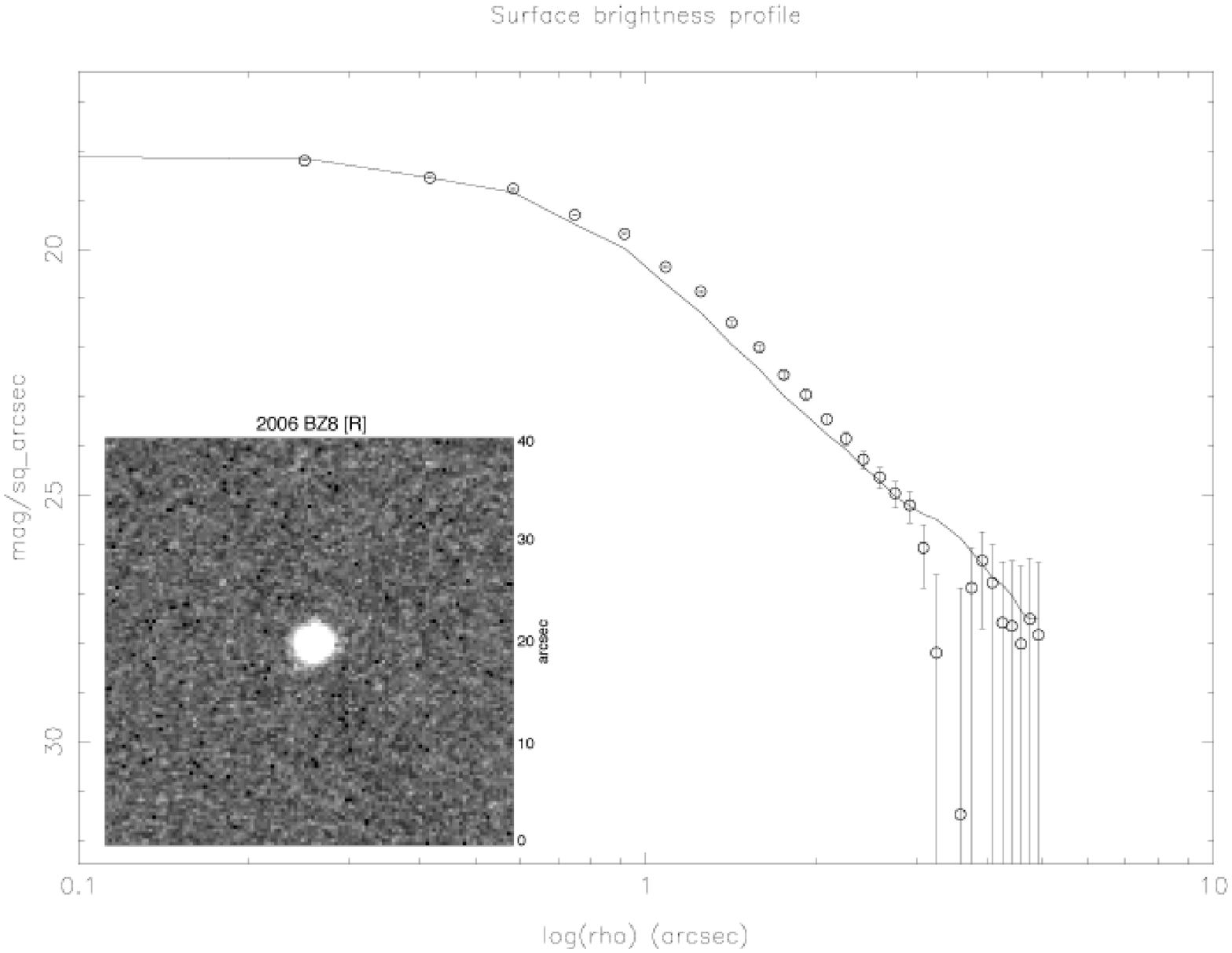} 
   \end{tabular}
     \caption[Images and SBP]{Images and surface brightness profiles for unresolved snap-shot targets. From the top left, these are 8P, 70P, 114P, 120P, 131P, 160P, P/1995~A1, P/2004~H3, and the Damocloid 2006~BZ8. Notes: {\it 70P} -- This is a single 295 s $r'$-band frame; in subsequent frames the comet moved towards the two stars and was not separable from them. The profile shows the effect of the nearby star; no sensible constraints on the activity level of this comet can be made. {\it 114P} -- taken on 4/7/05, a sum of 8$\times$ 90 s $r'$-band frames. The background stars are trailed in this image as it has been corrected for the comet's motion; with only 8 frames it was not possible to construct a good background image to subtract the stars. The profile is inactive in the inner part, although dominated by sky noise and the star trails in the outer section. {\it 120P} -- taken on 2/3/06, which is seen to be very faint, but distinguishable from the trailed stars in this combination of 4 $r'$-band frames. The star trails give two additional peaks in the profile at larger $\rho$, although the inner 1\arcsec{} implies a bare nucleus. {\it 131P} -- as a faint but star-like source in a combination of 3 frames taken on 2/3/06. The profile is in agreement with the image PSF within the error bars, implying that the comet was inactive. {\it 160P} -- which was detected at a very faint level, even in this combined image of three 140 s $r'$-band frames taken on 2/3/06. The profile appears stellar in the inner part, but is confused with a background galaxy in the outer part. {\it P/1995~A1} -- which was recovered on 2/7/05 returning to the inner solar system 10 years after it was last observed. The orbit is accurate, as the predicted motions allowed the 8 frames to be combined to give a star-like comet image (with trailed background stars). The profile matches the PSF, even in the outer noisy part, except at $\rho \approx 5\arcsec$ where flux from the nearby star trail raises the measured surface brightness.  {\it P/2004~H3} -- taken on the 5th of July 2005, in which the comet is just detectable as a very faint point source among the trailed stars. The profile is noisy due to the extreme faintness of the comet, but is consistent with an inactive nucleus. {\it 2006~BZ8} -- This Damocloid asteroid was observed on 2/3/06 and was bright in individual 10 s exposures, and is clearly stellar in this combination of all 15 $r'$-band exposures. The profile confirms this.
            }
         \label{unresolved_fig}
  \end{figure*}

\begin{figure}
   \begin{tabular}{c c}
   \includegraphics[width=0.24\textwidth]{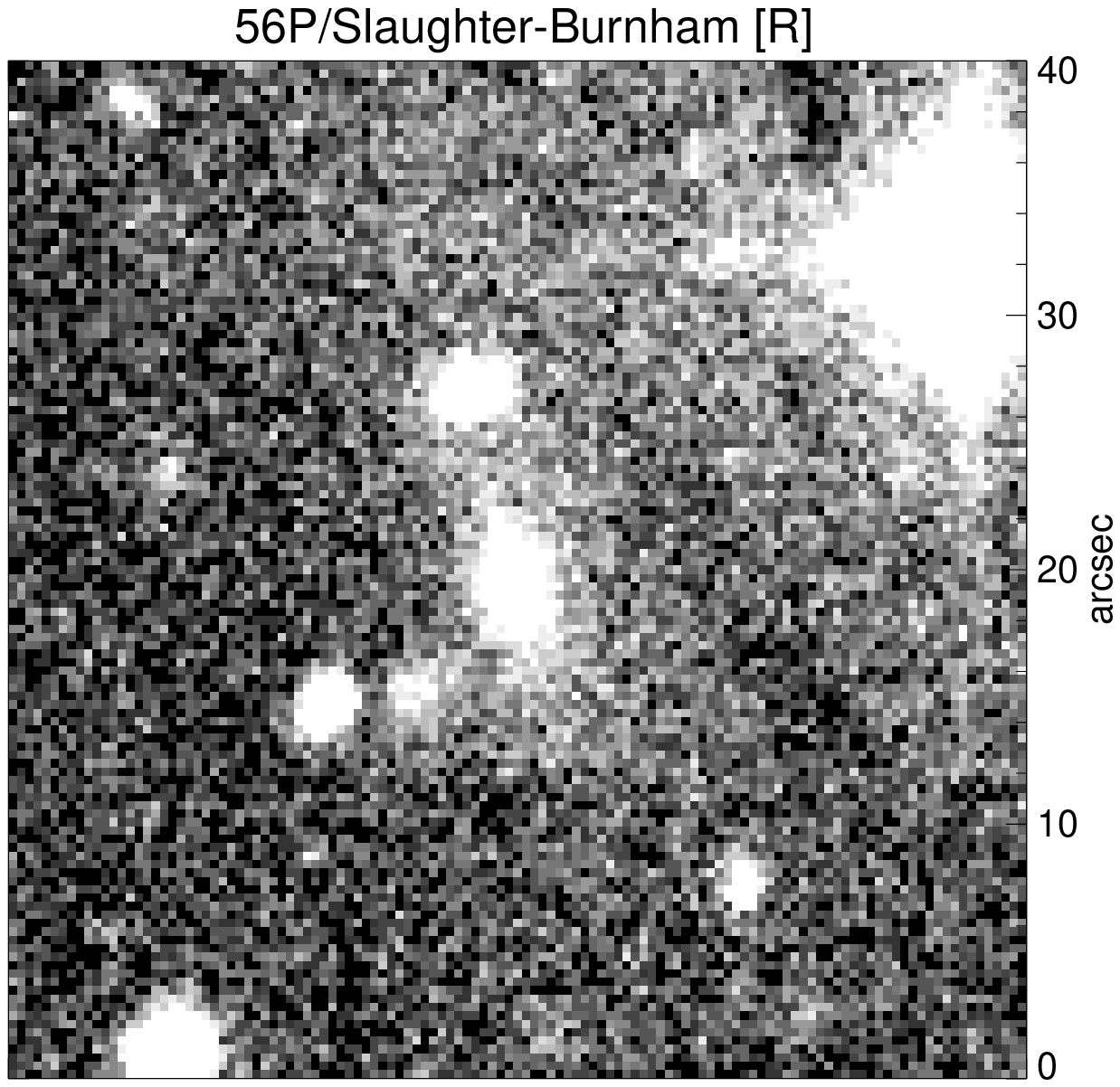} &
   \includegraphics[origin=br,angle=-90,width=0.2\textwidth]{56Psbp.ps} \\
  \includegraphics[width=0.24\textwidth]{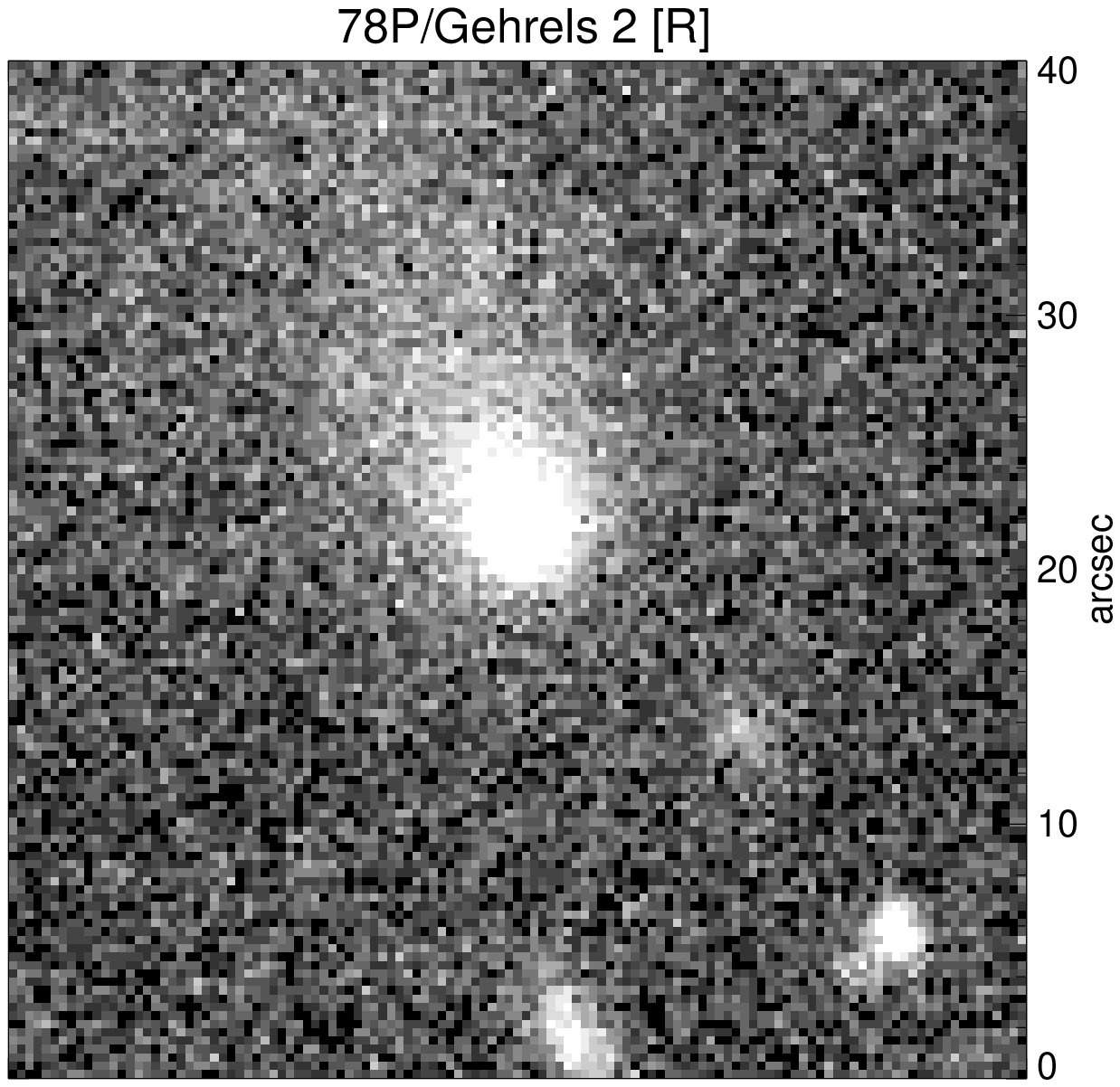} &
   \includegraphics[origin=br,angle=-90,width=0.2\textwidth]{78Psbp.ps} \\
   \includegraphics[width=0.24\textwidth]{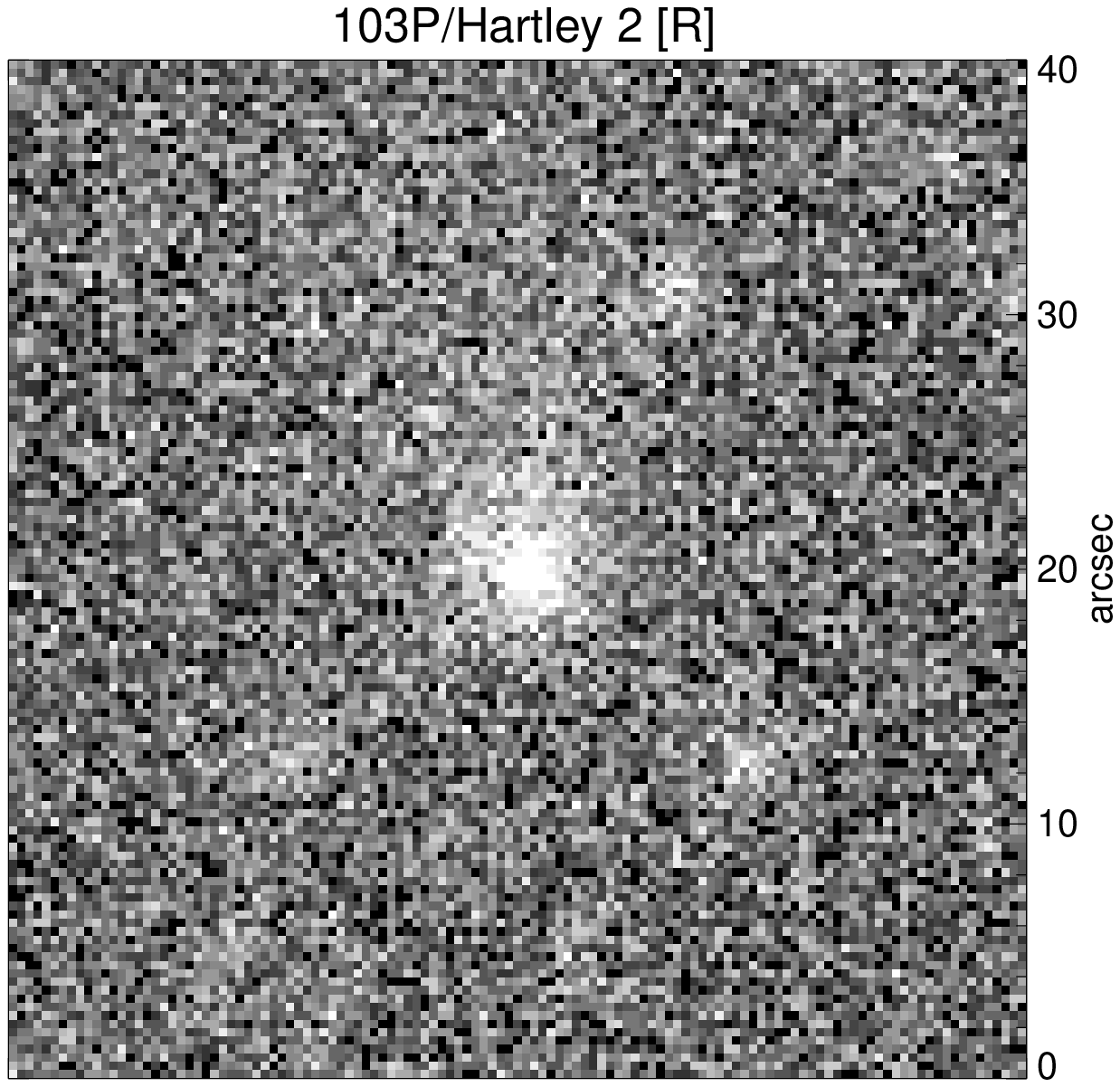} &
   \includegraphics[origin=br,angle=-90,width=0.2\textwidth]{103PsbpINT.ps} \\
   \end{tabular}
      \caption[Active comets images and profiles]{Images and surface brightness profiles for 56P, 78P and 103P. The comets are visibly active. The solid diagonal lines in the top right corner of each profile show gradients of -1 and -1.5, from two theoretical models of steady state coma; the comets all have profiles with gradients~$\sim -1.5$.}
         \label{active_fig}
   \end{figure}

For those comets which were detected, a surface brightness profile is measured to search for faint coma or quantify the level of activity. Images and profiles are shown in figures \ref{unresolved_fig} and \ref{active_fig}. Any coma present is assumed to be in a steady state, where the surface brightness is inversely proportional to the distance from the image centre $\rho$ (arcsec). For apparently inactive comets, we use the equation given by \citet{Jewitt+Danielson84}
\begin{equation}\label{sbp_eqn}
m_c(\rho) = \Sigma_c(\rho) - 2.5\log(2\pi\rho^2).
\end{equation}
to give the flux from the coma $m_c$ within $\rho$ based on the surface brightness $\Sigma_c$ at this distance. The fraction of the flux from any faint coma can then be found by comparing $m_c$ with the total flux within $\rho$. When the fraction is low, it can be assumed that the comet is effectively inactive, and that the nucleus has been observed. For active comets we make the usual $Af\rho$ measurement (\citealt{Ahearn84}; see also paper II), and place limits on the nucleus.

We obtain the radius of the nucleus $r_{\rm N}$ (or upper limits on it) based on the observed magnitude $m_R$ using \citep{Russell}
\begin{equation}\label{rneqn}
p_R r^2_{\mathrm{N}} = 2.238 \times 10^{22} R^2_{\rm h} \Delta^2 10^{0.4(m_\odot - m_R + \beta\alpha)}
\end{equation}
or, in terms of the absolute magnitude $m_R(1,1,0)$,
\begin{equation}
p_R r^2_{\mathrm{N}} = 2.238 \times 10^{22} 10^{0.4(m_\odot - m_R(1,1,0))}
\end{equation}
where $p_R$ is the geometric albedo, $R_{\rm h}$ and $\Delta$ the heliocentric and geocentric distances in AU, $m_\odot=-27.09$ is the apparent $R$-band magnitude of the Sun, $\alpha$ is the phase angle in degrees and $\beta$ is a linear phase co-efficient in magnitudes per degree. Following our previous papers, we assume $p_R$ = 4\% and $\beta$=0.035 for these unknown parameters.


\subsection{Undetected comets}

\subsubsection{72P/Denning-Fujikawa}
The radius of comet 72P/Denning-Fujikawa was unknown before the observations were obtained therefore we could not reliably predict its brightness. Five $r'$-band exposures were taken, giving a total exposure time of 250 seconds, but the comet was not detected, even in a combined image. Historically, this comet has proved elusive: It was first discovered in October 1881, but lost until a chance recovery during a close pass to the Earth in 1978, which was linked to the earlier comet despite it being unobserved during any of the 10 intervening apparitions\footnote{The history of previous observations of comets are taken from Gary Kronk's Cometography ({\tt http://cometography.com/}). Dates of last observation and orbital parameters are also sourced from the Minor Planets Center ({\tt http://cfa-www.harvard.edu/iau/mpc.html}) and the JPL small bodies database ({\tt http://ssd.jpl.nasa.gov/sbdb.cgi})}. Subsequent attempts to recover the comet during its 1987, 1996 and 2005 apparitions have failed; the non-detection reported here continues this trend. \citet{Lamy-chapter} quote an unpublished radius estimate from the updated catalogue of Tancredi~et~al.~of 0.8 km, but it is far from clear what data this estimate is based on. If it were true, then this radius implies a predicted $m_R = 22.4$ at the time of the INT run, which should have been just detectable. 

There are three likely interpretations of this non-detection; firstly, the orbit could be incorrect, and in fact Denning in 1881 and Fujikawa in 1978 discovered two different comets. Assuming that this is not the case, the alternative explanations are that the comet has ceased to exist ({\it i.~e.}~broken up) since 1978, or that it is a very small nucleus that exhibits little out-gassing near perihelion, and has consequently escaped detection despite numerous attempts. If the comet exists, and was in the predicted position, then the fact that no object was observed to $m_R \ge 23.0$ implies an absolute magnitude for this comet of $m_R(1,1,0) \ge 18.3$ and therefore $r_{\rm N} \le 0.6$ km.

\subsubsection{75P/Kohoutek and P/2001~X2 (Scotti)}

Both 75P/Kohoutek and P/2001~X2 (Scotti) were observed during the 2005 INT run at right ascension 18 h - 19 h and declination $\sim$ -20\degr, which unfortunately placed them in the galactic plane [galactic co-ordinates $(l,b) \approx$ (10,0) - (20,-10)]. This meant that both were observed against a dense star field, making identification of the comets difficult. For P/2001~X2 this was exacerbated by the loss of one of the frames taken due to a malfunction in the CCD camera, leaving only two $r'$-band frames. It proved impossible to identify either comet, even with the aid of astrometric predictions of their precise positions within these fields. As there is a high likelihood of the comet being blended with one of the stars within the predicted area it is also impossible to constrain the brightness of the comets other than to say that they must be considerably fainter than the field stars, but this does not put meaningful limits on the radius as for any sensible size the comet would fit this constraint: for 75P the stars in the search area have magnitudes in the range $m_R \approx 14.5 - 20.5$, while the comet is expected to have $m_R \ge 22.8$ based on a limit of $r_{\rm N} \le 1.6$ km \citep{Lowry03}\footnote{Note that the $R$-band magnitude of the Sun used by \citet{Lowry99,Lowry03} and \citet{Lowry+Fitzsimmons01,Lowry+Fitzsimmons05}, $m_\odot = -27.26$, was based on the often quoted value of $(V-R)_\odot=0.52$ from \citet{Allen}, which is for the Johnson filter system, not the modern Johnson-Cousins system used by \citet{Landolt}. In the Johnson-Cousins system $(V-R)_\odot=0.35$ \citep*{Holmberg06}. The difference between $m_\odot = -27.26$ and the true value of $m_\odot = -27.09$ means that while the magnitudes measured in these works are accurate, the true radii are $10^{0.2(-27.09 + 27.26)} = 8\%$ larger. The radius for 75P, and for all others quoted from this group of papers, have been corrected to account for this.}, from a non-detection in 1999. 

\subsubsection{135P/Shoemaker-Levy~8}

135P/Shoemaker-Levy~8 was inbound at $R_{\rm h}$ = 3.6 AU at the time of the 2006 INT run, on its way to perihelion at 2.7 AU in May 2007, although it has yet to be recovered (as of 11 July 2007). The comet was not detected in any of the five 80 second $r'$-band exposures or in a combined frame.  There is no object at the position given by an astrometric fit to $m_R \ge 23.5$, giving a limit to the absolute magnitude of $m_R(1,1,0) \ge 18.4$ and therefore a radius of $r_{\rm N} \le 0.6$ km.

\subsubsection{P/2004~T1 (LINEAR-NEAT)}

P/2004~T1 (LINEAR-NEAT) was discovered in September 2004 before passing through perihelion at 1.7 AU in November 2004. It was last observed outbound in March 2005, and was still outbound at 3.8 AU during February/March 2006. No further information has been recorded on this comet. 5$\times$60 second $r'$-band exposures were taken, but the comet was not detected. An astrometric fit and measurement of a limiting magnitude of $m_R \ge 23.6$, imply $m_R(1,1,0) \ge 18.3$ and $r_{\rm N} \le 0.6$ km.


\subsection{Unresolved comets}

\subsubsection{8P/Tuttle}

The results from the snap-shot observations of 8P/Tuttle obtained during the July 2005 INT run are presented here, although they are not included in the analysis of the ensemble properties of ecliptic comets. 8P is an unusual comet; by the `classical' definition, it is a JFC, as it has an orbital period less than 20 years ($P = 13.5$ yr), however on dynamical grounds it is a Nearly Isotropic Comet \citep[see][]{Levison96}, as it has a Tisserand parameter with respect to Jupiter of $T_J = 1.6$, and has a Halley-like high inclination ($i=55\degr$) orbit. It therefore is likely to be an Oort cloud comet which has been captured into a shorter period orbit by interactions with the planets. It is also known to have a reasonably large ($r_{\rm N}$ = 7.8 km) nucleus, from snap-shot observations by \citet{Licandro00b}, and was therefore predicted to be a bright target suitable for time-series observation at $R_{\rm h}$ = 7.4 AU with the INT, providing an opportunity to add to the very few data on the rotational properties of Oort cloud comets. However, upon observation the target field was seen to be far too crowded for time-series observation, and attempts to observe the comet during this run were abandoned after only a short series of 13 $r'$-band frames. It was possible to separate the comet from background sources in the majority of frames; the average $m_R = 21.61\pm0.03$ is taken over the 9 of these frames over which the magnitude of the comet was $\sim$ constant, with a brightening in the remaining 4 being attributed to proximity to a background star. This magnitude gives $m_R(1,1,0) = 12.88$, corresponding to a radius of $7.58\pm0.12$ km, in excellent agreement with the value of \citet{Licandro00b}. 

\subsubsection{44P/Reinmuth~2}

A snap-shot of 44P/Reinmuth~2 was published in paper II. It was observed again during the 2006 INT run, when a total of 5 $r'$-band images were taken over the course of the two nights. The comet was detected on both nights, although was very faint in poor seeing on the first. The comet's magnitude was constant, within the error bars on each point, over the two nights at $m_R = 22.49\pm0.06$, and apparently inactive. The profile is consistent with the background PSF, within the error bars on each point, and equation \ref{sbp_eqn} gives $m_c \ge 23.26$, or $\le 50\pm72\%$ of the flux. The absolute magnitude was measured to be $m_R(1,1,0) = 15.95$, implying $r_{\rm N}=1.84\pm0.05$ km, which is consistent with our previous measurement of $r_{\rm N}=1.96\pm0.11$ km. Colours of $(V-R)  = 0.31\pm0.09$ and $(R-I)  = 0.42\pm0.13$ were measured.

\subsubsection{70P/Kojima}

\citet{Lamy-chapter} quote measurements for 70P/Kojima from their own unpublished work, from a `partial rotational light-curve', although no indication is given of the quantity of data involved. Accordingly, their results ($r_{\rm N}$ = 1.86 km, $a/b \ge 1.1$ and $P_{\rm rot} \ge 22$ hours) are treated with caution, although the radius measurement can be taken at face value if regarded as a snap-shot. Identification of the comet at the telescope was hampered by the faintness of the target and its proximity to a bright field star; only in the first frame was it sufficiently separated from the star to allow photometry. Even in this first frame a small amount of flux from the star may have been present within the aperture centred on the nucleus; the following radius should be considered to be an upper limit. The measured comet magnitude is $m_R = 22.53\pm0.15$, which corresponds to $m_R(1,1,0) = 15.59$ and $r_{\rm N} \le 2.18 \pm 0.15$ km. Under the assumption that this is unaffected by star light, and taken with the radius measurement from \citeauthor{Lamy-chapter}, this implies $a/b \ge 1.4$, which is typical for JFCs. This is larger than \citeauthor{Lamy-chapter}'s value, but even if they did observe the full range in magnitudes in their partial light-curve then a change of viewing angle could increase the range observed at a different epoch: light-curves only give a {\em minimum} $a/b$ when the orientation of the rotation axis is unknown. Based on both data sets, an average $r_{\rm N} = 2.0\pm0.2$ seems a reasonable value to assume for the size of 70P.

\subsubsection{114P/Wiseman-Skiff}

114P/Wiseman-Skiff was observed with the INT in July 2005 when inbound at 3.75 AU, and was detected but at too faint a level to produce a reliable light-curve. The combined image of all 8 frames and corresponding profile show that the comet appeared inactive. The profile was only measurable to $\rho=4\arcsec$, but gives $m_c=24.45$ and therefore a limit on the coma contribution of $\le 25\pm41\%$, with the large error bars due to the noisy profile in the outer part. The measured $m_R = 22.95\pm0.11$ gives $m_R(1,1,0) = 17.34$ and $r_{\rm N} = 0.97\pm0.05$ km. This agrees with the assessment that the nucleus is a relatively small one from \citet{Lamy-chapter}, who again quote their own unpublished result of $r_{\rm N} = 0.78\pm0.04$, based on an HST snap-shot when the comet was at \hbox{$R_{\rm h}$ = 1.6 AU} and presumably active. Assuming that these two snap-shots were taken at opposing light-curve extrema, then the minimum required axial-ratio is $a/b \ge 1.5$, and the average effective radius is $r_{\rm N}$ = 0.89 km.

\subsubsection{120P/Mueller~1}

Although not detectable in individual images, 120P/Mueller~1 was found in a co-added image of 4$\times$ 115 s $r'$-band frames shifted to account for its predicted motion. The surface brightness profile for this comet matches the stellar PSF within the inner arcsecond, although beyond this there are two further peaks due to the nearby trails of background stars, making the formal limit of $m_c \ge 23.1$ meaningless. The comet had $m_R = 23.52\pm0.13$ in the combined image. This gives $m_R(1,1,0) = 17.84$ and $r_{\rm N} = 0.77\pm0.05$ km. Assuming that this measurement is of the radius of the bare nucleus, it can be compared with the previous measurement of \citet{Lowry99}. They measured $r_{\rm N}=1.7\pm0.2$ with a snap-shot of a star-like nucleus at $R_{\rm h}$ = 3.1 AU, which at approximately twice the radius measured here is a large discrepancy; such disparate measurements would require $a/b = 4.7$ to be explained purely by the shape of the nucleus, which is rather high. It is possible that there was either a non-zero coma contribution in \citeauthor{Lowry99}'s $R_{\rm h}$ = 3.1 AU measurement which was not present at 3.9 AU, or that the comet has a steeper than normal phase function / opposition surge (\citeauthor{Lowry99}'s observations were taken at $\alpha = 2.9\degr$). The unknown colour of the nucleus could also contribute, but only at a level similar to the uncertainty on $m_R$: If 120P has a very red nucleus the difference in colour term between \citeauthor{Lowry99}'s $R$-band and our $r'$-band measurements could cause the $m_R$ presented here to be up to a tenth of a magnitude too faint. It is likely that some combination of these factors, combined with a more modest elongation of the nucleus, is required to explain the differences between these measurements.

\subsubsection{131P/Mueller~2}

131P/Mueller~2 was outbound at 3.5 AU when observed with the INT in 2006. Three $r'$-band frames were taken, and the comet was found and observed to have a faint, star-like appearance. The large error bars on the surface brightness at large $\rho$ make the coma limit of $m_c \ge 23.1$ ($\le 68\pm94\%$ of the total flux) almost meaningless, but the profile appears star-like and the comet can be assumed to have been inactive. The average magnitude was $m_R = 22.77\pm0.09$, corresponding to an absolute magnitude of $m_R(1,1,0) = 17.59$, and therefore a radius of $r_{\rm N} = 0.87\pm0.04$ km. In addition, three $V$-band frames were taken, allowing measurement of the colour of the nucleus. A value of $(V-R)  = 0.45\pm0.12$ was obtained, which falls in the centre of the observed JFC distribution (see section \ref{discussion}).

\subsubsection{160P/LINEAR}

Observations of 160P/LINEAR from the INT during 2006 detected the comet when outbound at $\sim 4$ AU. The comet appeared as a very faint but apparently inactive object in 3$\times$140 s $r'$-band exposures. The presence of a nearby galaxy prevents measurement of limits on any coma, with $m_c(5\arcsec) = 22.6$. The inner part of the profile is a good match for the background PSF; measuring the surface brightness at 1.3\arcsec{} gives $m_c \ge 28.0$, $\le 1.9\pm5.8 \%$ of the total flux. The measured value of $m_R = 23.69\pm0.18$ implies $m_R(1,1,0) = 17.58$ and $r_{\rm N} = 0.87\pm0.07$ km. 

\subsubsection{P/1995~A1 (Jedicke)}

An attempt to detect the return of P/1995~A1 (Jedicke) was made in July 2005 with the INT, although the comet was still at $R_{\rm h}$  = 5.5 AU. A faint object was discovered near the nominal position with an average $m_R = 22.63\pm0.07$. Measured in a combination of all 8 $r'$-band frames, the comet's surface brightness profile appeared stellar in the inner region, although was dominated by sky noise beyond 2\arcsec, which is taken to show that it was inactive, as would be expected with the comet still beyond Jupiter's orbit. This implies a radius for the nucleus of $r_{\rm N} = 2.48\pm0.08$, and an absolute magnitude of $m_R(1,1,0) = 15.31$.

\subsubsection{P/2004~H3 (Larsen)}

P/2004~H3 (Larsen) was only at its second opposition when observed with the INT in July 2005, having been discovered the previous April. It was at $R_{\rm h}$ = 3.7 AU, outbound, and was only just recovered in a frame combining all six 90 s $r'$-band exposures. The measured magnitude, in the combined frame, was $m_R = 24.30\pm0.21$, implying that the nucleus is small, with $m_R(1,1,0) = 18.67$ and a corresponding radius of $r_{\rm N} = 0.53\pm0.05$ km. A profile is just measurable, but only within 3\arcsec, and is very noisy. Formally, the coma magnitude from equation \ref{sbp_eqn} is $m_c \ge 24.6$, implying a flux contribution of $\le 80\%$, although the error on this is also $\sim 80\%$, making it of limited value. The profile matches the stellar PSF at all radii, within the error bars.

\subsubsection{2006~BZ8}

Observations of 2006~BZ8 were carried out to search for activity around this newly discovered object. It was regarded as being almost certainly a comet, with $T_J = -1.0$. No activity was observed though, and the object was bright ($m_R = 16.858\pm0.003$) and stellar. A surface brightness profile entirely matched the combined frame PSF, and equation \ref{sbp_eqn} gave a limit on any coma at $m_c \ge 21.67$, or $\le 1.2\pm1.2 \%$ of the flux. 2006~BZ8 is officially a Centaur\footnote{Objects with $5.5 \le a \le 30.1$ AU are designated Centaurs.}, as it has an orbital semi-major axis of $a=9.7$ AU, however it seems likely that it is from the Oort cloud due to its Tisserand parameter and high inclination ($i = 165\degr$), and is better classified as a member of the Damocloid family \citep{Jewitt05}. The absolute magnitude was measured to be $m_R(1,1,0) = 14.14$, corresponding to $r_{\rm N} = 4.22\pm0.02$ km, and its colours were found to be $(V-R)  = 0.62\pm0.01$ and $(R-I)  = 0.36\pm0.01$. These results are not included in the discussion on the ensemble properties of JFCs in section \ref{discussion}.


\subsection{Active comets}

\subsubsection{56P/Slaughter-Burnham}

56P/Slaughter-Burnham was discovered in 1958, and has previously been found to have $r_{\rm N}$ = 1.56 km \citep{Meech04}. This led to a predicted nuclear magnitude of $m_R = 22.1$ at the time of the 2006 INT observations, and the comet was well placed for time-series observations at the beginning of the night. The comet was active at $R_{\rm h}$ = 3.8 AU (outbound). The mean magnitude was $m_R = 20.73\pm0.04$, implying $m_R(1,1,0) \ge 14.90$ and $r_{\rm N} \le 3.00$ km, in broad agreement with the measurement by \citet{Meech04}. The colours of the comet were measured to be $(V-R)  = 0.51\pm0.05$ and $(R-I)  = 0.35\pm0.08$, which fall in the observed range for nuclei, although it should be noted that these colours represent the colours of the coma, as the comet was active. These colours are similar to other measurements of the broad-band colours of dust comae \citep{Lowry99,Lowry03}, and also appear to be in agreement with the trend of cometary dust to have more neutral (less red) colours at longer wavelengths, which is interpreted as showing that the dominant grains are larger than 1 \micron{} in size \citep{Jewitt+Meech86}. It cannot be assumed that the coma has the same colour as the surface of the nucleus for highly active comets; even if the dust component is of similar composition, there is likely to be gas present also, especially in the $V$-band. The flux within an aperture of $\rho = 5\arcsec (\equiv 11400$ km at the comet) gives $Af\rho = 22.3\pm0.3$ cm.

\subsubsection{78P/Gehrels~2}

Three $r'$-band frames were taken of 78P/Gehrels~2 during the 2006 INT run, which showed the comet to be clearly active when outbound at $R_{\rm h}$ = 3.8 AU. \citet{Lowry+Weissman03} measured a radius of $r_{\rm N} = 1.54\pm0.12$ km from a snap-shot taken when the comet was at $R_{\rm h}$ = 5.5 AU; the upper limit obtained here of $r_{\rm N} \le 4.21$ km, from $m_R = 19.42\pm0.01$ and $m_R(1,1,0) \ge 14.16$, is consistent with this. The activity level, measured within $\rho = 5\arcsec (\equiv 10400$ km), was found to be $Af\rho = 18.2\pm0.2$ cm.

\subsubsection{103P/Hartley~2}\label{103Psect}

103P/Hartley~2 was observed to be highly active during the 2005 NTT run (see paper II). It was observed again during the 2006 INT run, despite a strong suspicion that it may still be active at $R_{\rm h}$ = 5.0 AU, as at the time this comet was a potential target for the {\it Deep Impact} extended mission, {\it EPOXI}\footnote{{\tt http://epoxi.umd.edu/} -- 103P has now been confirmed as the target of this mission, with a fly-by date in 2010.}. The comet was detected in two $r'$-band frames taken on the first night, and was clearly active. The measured $m_R = 21.00\pm0.05$ implies a radius upper limit of $r_{\rm N} \le 4.45$ km (it is thought that the nucleus is actually much smaller; see paper II), while the flux within $\rho = 15600$ km gives $Af\rho = 47.2\pm0.7$ cm. The coma had colours of $(V-R)  = 0.16\pm0.09$ and $(R-I)  = 0.35\pm0.08$. The rather blue $(V-R)$  measurement could reflect the presence of significant flux from gas emissions within the $V$-band, such as the C$_2$ bands at 5165 and 5635 \AA, while the $(R-I)$  colour is similar to that found for nuclei and dust comae.

\begin{table*}
\begin{minipage}[]{\textwidth}
\caption[Snap-shot results]{Derived values and limits on radii and activity from snap-shot photometry.}             
\label{results_snapshots}      
\renewcommand{\footnoterule}{}  
\renewcommand{\arraystretch}{1.5}
\begin{tabular}{l c c c c c}        
\hline\hline                 
Comet & $m_R$ & $m_c$\footnote{Coma magnitude within $\rho =5\arcsec{}$ (unless stated otherwise) measured using equation \ref{sbp_eqn}.} & $m_R(1,1,0)$ & $r_{\mathrm{N}}$ [km] & $Af\rho$ [cm]\footnote{$Af\rho$ measured through an aperture of radius 5\arcsec.}  \\    
\hline
\multicolumn{3}{l}{\bf UNDETECTED}\\
72P & $\ge$ 23.0  & - & $\ge$ 18.3 & $\le$ 0.6 & $\le$ 0.45 \\
75P & - & - & - & - & - \\
135P & $\ge$ 23.5 & - & $\ge$ 18.4 & $\le$ 0.6 & $\le$ 0.88  \\
P/2001 X2 & - & - & - & - & -  \\
P/2004 T1 & $\ge$ 23.6 & - & $\ge$ 18.3 & $\le$ 0.6 & $\le$ 0.57  \\
\hline
\multicolumn{3}{l}{\bf UNRESOLVED}\\
8P & 21.61$\pm$0.03 & $\ge$25.1 (4.3\arcsec) & 12.88$\pm$0.03 & 7.58$\pm$0.12 & - \\
44P & 22.49$\pm$0.06 & $\ge$23.3 & 15.95$\pm$0.06 & 1.84$\pm$0.05 & $\le$1.7 \\
70P & 22.53$\pm$0.15 & $\ge$23.9 & 15.59$\pm$0.16 & 2.18$\pm$0.15 & $\le$23.7\\
114P & 22.95$\pm$0.11 & $\ge$24.5 (4.0\arcsec) & 17.34$\pm$0.11 & 0.97$\pm$0.05 & $\le$0.24\\
120P & 23.52$\pm$0.13 & $\ge$23.1 & 17.84$\pm$0.13 & 0.77$\pm$0.05 & $\le$2.9\\
131P & 22.77$\pm$0.10 & $\ge$23.1 & 17.49$\pm$0.10 & 0.87$\pm$0.04 & $\le$0.50\\
160P & 23.69$\pm$0.18 & $\ge$28.0 (1.3\arcsec) & 17.58$\pm$0.18 & 0.87$\pm$0.07 & $\le$4.7\\
P/1995 A1 & 22.63$\pm$0.07 & $\ge$22.1 & 15.31$\pm$0.07 & 2.48$\pm$0.08 & $\le$5.7\\
P/2004 H3 & 24.19$\pm$0.21 & $\ge$24.6 (3.0\arcsec) & 18.55$\pm$0.21 & 0.56$\pm$0.06 & $\le$0.69\\
2006 BZ8 & 16.88$\pm$0.01 & $\ge$21.7 & 14.16$\pm$0.01 & 4.22$\pm$0.02 & $\le$41.4\\
\hline
\multicolumn{3}{l}{\bf ACTIVE}\\
56P & 20.73$\pm$0.04 & - & $\ge$14.90 & $\le$3.00 & 22.2$\pm$0.3\\
78P & 19.42$\pm$0.01 & - & $\ge$14.16 & $\le$4.21 & 18.2$\pm$0.2\\
103P& 21.00$\pm$0.05 & - & $\ge$14.04 & $\le$4.45 & 47.2$\pm$0.7 \\
\hline                                   
\end{tabular}
\end{minipage}
\end{table*}


\section{Time-series photometry}
\label{lc-results}

Those comets which appeared (upon visual inspection at the telescope) to be inactive and reasonably bright were followed over an extended period in order to obtain a light-curve. We assume that the variation in brightness is due to the nucleus being a rotating non-spherical body, and that the variation in reflected flux is due to the varying projected surface area. Here we follow the common practice of describing the nucleus as the simplest non-spherical shape: a tri-axial ellipsoid with semi-axes $a \ge b = c$. We therefore expect to see a double peaked light-curve with a rotation period $P_{\mathrm{rot}}$ twice the fitted Fourier period $P_{\mathrm{fitted}}$. The Fourier method used to search for periodicities is described in paper I; briefly, we search for minima in the range $\chi^2/\nu = 1 \pm \sqrt{2/\nu}$ in a reduced $\chi^2$ periodogram. $\nu$ is the number of degrees of freedom of the model, given by $\nu = (N - 3)$ where $N$ is the number of data points, for a first order Fourier model.

The periodicity search allows us to constrain the rotation period, while the peak-to-trough amplitude of the light-curve $\Delta m$ gives the minimum elongation of the nucleus via $a/b \ge 10^{0.4\Delta m}$. We can only measure a lower limit on $a/b$, as we do not know the orientation of the rotation axis. We use the approximation given by \citet{Pravec+Harris00};
\begin{equation}\label{densityeqn}
D_{\mathrm{N}} \ge \frac{10.9}{P^2_\mathrm{rot}} \frac{a}{b},
\end{equation}
to calculate the minimum bulk density of each nucleus using the light-curve derived parameters $a/b$ and $P_{\rm rot}$ (hours). Note that this density value is a minimum for two reasons: it relies on $a/b$, for which we have only a lower limit, and also that the nucleus need not be spinning at its break up rate, but clearly cannot be spinning faster than it. The results from the time-series observations are summarised in table \ref{results_lightcurve}.

\subsection{40P/V{\" a}is{\" a}l{\" a}~1}

\begin{figure}
   \includegraphics[width=0.47\textwidth]{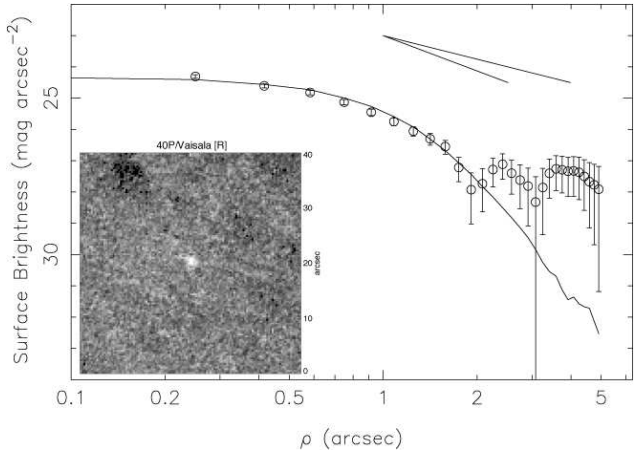}
      \caption[40P image \& SBP]{Image showing 40P, made up of 20$\times$85 second exposures taken on the 6th night of the 2005 INT run. The profile is inconclusive, but we suspect that the comet was weakly active.}
         \label{40Pimage}
   \end{figure}

40P/V{\" a}is{\" a}l{\" a}~1 was observed over 7 nights during the 2005 INT run. It had a mean $m_R = 22.08\pm0.02$. The surface brightness profile is inconclusive, but we suspect that the comet was weakly active at $R_{\rm h}$ = 4.6 AU. The coma contribution, assuming steady state, was measured to be $m_c \ge 22.22$ within 5\arcsec{}, and therefore up to $\sim 95\%$ of the flux, but the irregular shape of the profile (fig. \ref{40Pimage}) implies that the steady state assumption may not be valid in this case.

With observations over 7 nights, changes in the relative positions of the Earth and the comet meant that there was a non-negligible change in observing geometry. The change in apparent magnitude due to variations in $r_{\rm H}$, $\Delta$ and $\alpha$ between the first and last frames is $\delta m_R \approx 0.04$ mag., which although small compared with the uncertainty on individual points (typically $\sim0.2$ mag.) is considerable when compared to the error on the average magnitude. The photometry was therefore reduced to the appropriate $m_R(1,1,0)$ using the precise position of the comet at the time of each observation, which was generated using {\sl HORIZONS}\footnote{{\tt http://ssd.jpl.nasa.gov/horizons.cgi}}. The phase function is assumed to be linear with $\beta = 0.035$ mag.~deg$^{-1}$, as the range in phase angle over 7 nights is not large enough to independently measure $\beta$ for 40P. The mean $m_R(1,1,0) = 15.72$ gives an upper limit to the radius of $r_{\rm N} \le 2.05\pm0.02$ km, assuming a 4\% albedo. This is a stronger constraint than the previous upper limit of $r_{\rm N} \le 3.6$ km from \citet{Lowry99}, who did not detect the comet when it was at $R_{\rm h}$ = 6.01 AU, outbound, in 1995. If the coma contamination is large, as suggested above, then the true radius of the nucleus could be considerably smaller than this.

   \begin{figure}
   \includegraphics[origin=br,angle=-90,width=0.47\textwidth]{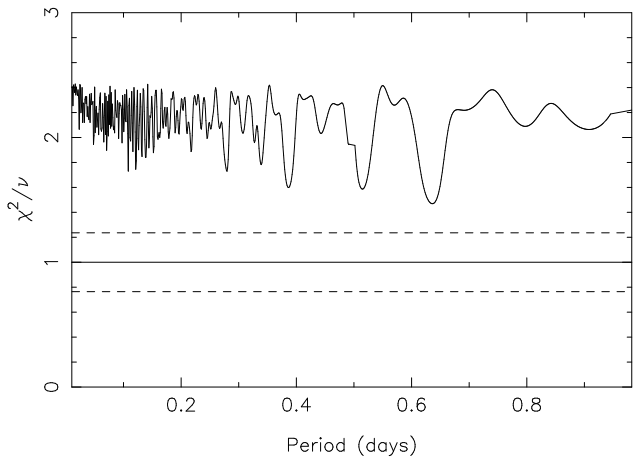}
      \caption[40P periodogram]{Periodogram for 40P. No convincing period was found.}
         \label{40Ppgram}
   \end{figure}

The absolute magnitude variations were searched for periodicities, but none were found which were convincing. The strongest minimum in the periodogram (fig. \ref{40Ppgram}) corresponds to $P_{\rm fitted} =15.3$ hours, and therefore a long rotation period of 30.6 hours, but does not give a visually acceptable light-curve at either of these periods. Periodogram searches were also carried out on short subsets of the data, using fully differential light-curves for each pair of nights, during which the comet's motion was small enough to give common comparison stars. Shorter periods in these data are neither particularly convincing in the subsets, due to sparse data, nor good fits to the full data set. Although there is a large range in observed magnitudes ($\Delta m \approx 1.4\pm0.3$ mag), this appears to be due to variations in the dust coma, and is not due to the nucleus, again implying that in this case the coma completely dominates. 

The colour measurements have large uncertainties due to the comet being very faint, but are consistent with means of $(V-R)  = 0.37 \pm 0.10$ and $(R-I)  = 0.65\pm0.11$. At large distance from the Sun the coma should be dominated by dust, with very little gas present; the colours measured for the comet could therefore be similar to the colours of the nucleus, although this is actually a measurement of the colour of the dust coma. 

\subsection{47P/Ashbrook-Jackson}

   \begin{figure}
 \begin{tabular}{c c}
   \includegraphics[width=0.24\textwidth]{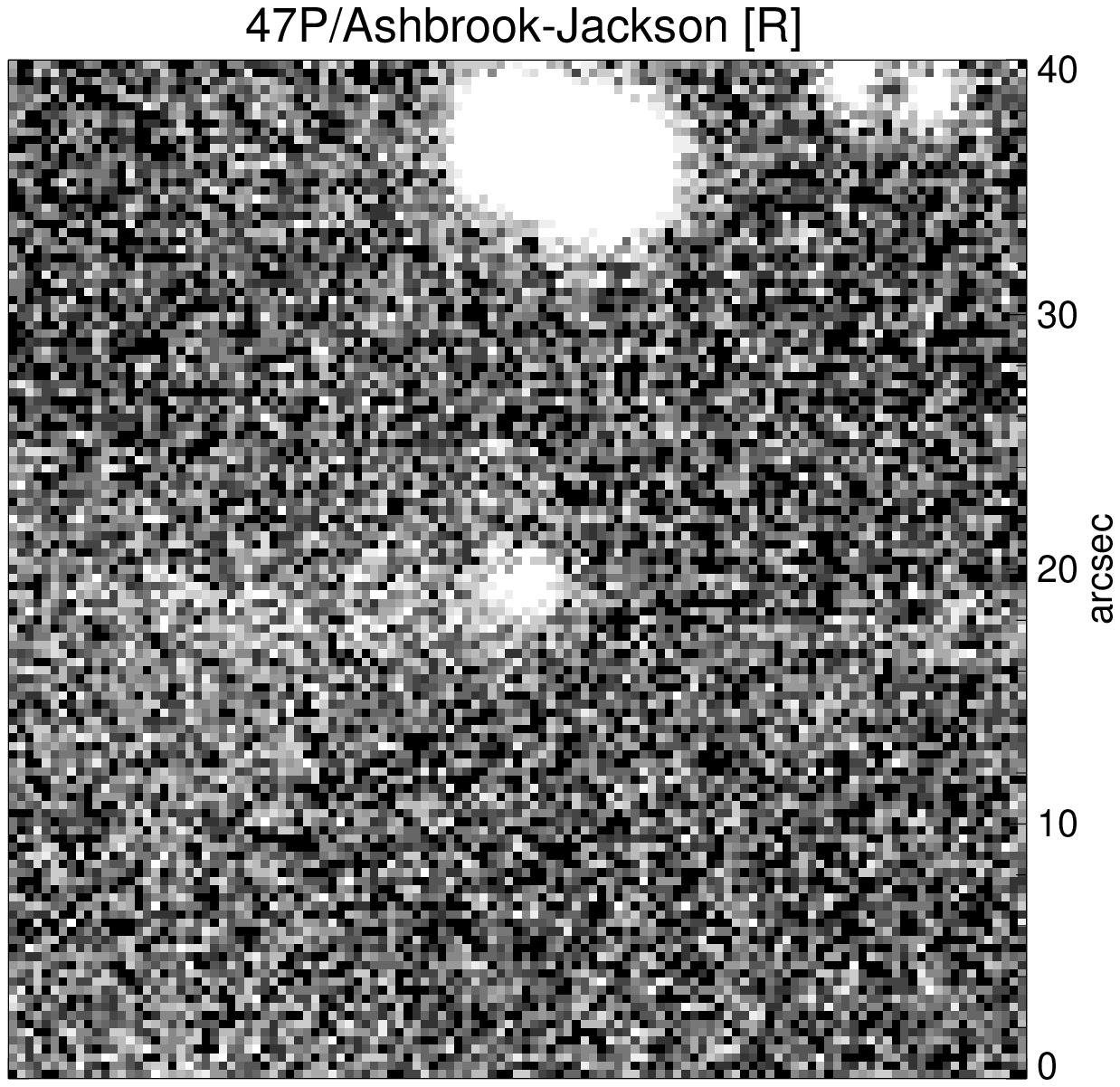}&
   \includegraphics[origin=br,angle=-90,width=0.2\textwidth]{47PsbpINT3.ps} \\
   \includegraphics[width=0.24\textwidth]{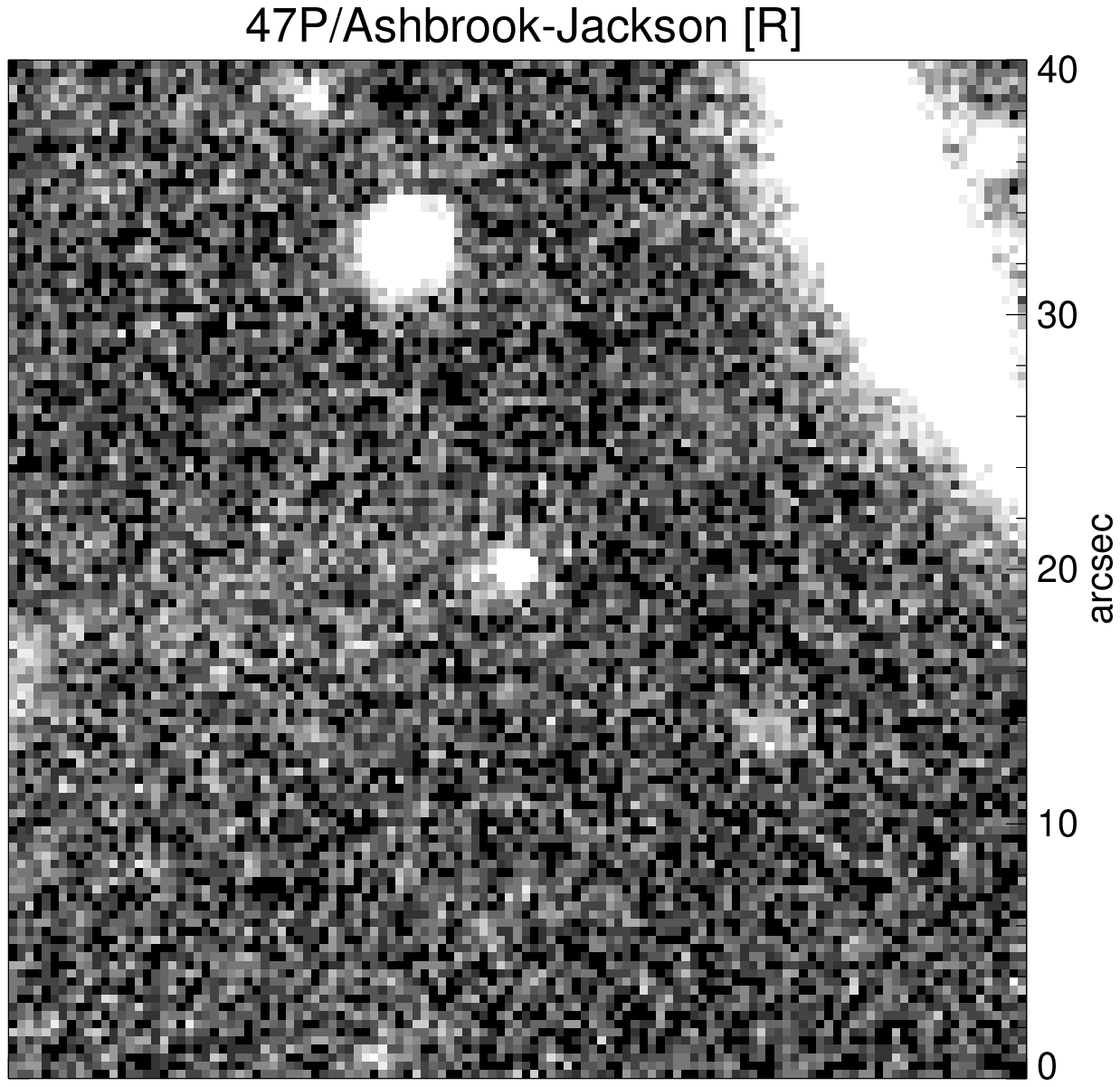}&
   \includegraphics[origin=br,angle=-90,width=0.2\textwidth]{47PsbpINT4.ps}
   \end{tabular}

      \caption[47P image \& SBP - INT]{Images of 47P taken during the 2006 INT run. There is a faint tail visible to the West (left in these images) of the comet, which is clearer in the night 3 data (top). The profile shows activity. In the night 4 data (bottom), the tail is less obvious, as the total exposure time is less.}
         \label{47PimageINT}
   \end{figure}

   \begin{figure}
   \includegraphics[origin=br,angle=-90,width=0.47\textwidth]{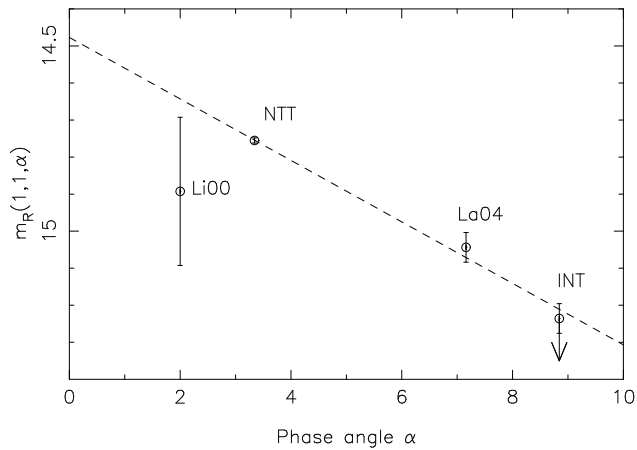}
      \caption[47P phase function]{47P mean magnitudes from paper II, the INT data and from the literature reduced to $r_{\rm H}=\Delta=1$ AU and plotted as a function of phase angle $\alpha$. There is a good fit through the time-series data giving a linear phase law with $\beta = 0.083$. Labels give the sources of each point; Li00: \citealt{Licandro00b}; La04: \citealt{Lamy-chapter}; NTT: paper II. The faint activity seen in the INT data means that it provides a lower limit on $m_R(1,1,\alpha)$, although the coma clearly contributes little flux as making this point any fainter would mean an even steeper phase function and a poor fit.}
         \label{47Pphase}
   \end{figure}

A time series on 47P/Ashbrook-Jackson was published in paper II. The comet was observed again during the 2006 INT run; a total of 5 $r'$-band exposures totalling 18.5 minutes were taken over the two nights. Upon processing these data, 47P was found to be faintly active, despite still being beyond 5 AU, with a faint tail visible to the West of the comet (fig. \ref{47PimageINT}). The activity is clearly weak: The profile measured in good ($\sim1\arcsec$) seeing on the last night matches the stellar PSF within the inner part, where the nucleus appears to dominate the flux. The activity level can be quantified as $Af\rho = 3 - 11$ cm, with the lower number being measured on the latter night. 

The average magnitude over the two nights is $m_R = 22.01\pm0.04$, with a total variation similar to that observed in the light-curves above, of $\Delta m = 0.4$ mag. These imply $m_R(1,1,0) \ge 14.93$ and $r_{\rm N} \le 2.96\pm0.05$ km, with $a/b \ge 1.4$, slightly fainter than our earlier results despite the weak coma, possibly due to a steeper phase function than $\beta=0.035$ mag.~deg$^{-1}$. Taking both the NTT and INT results, and those of \citet{Licandro00b} and \citeauthor{Lamy-chapter}\footnote{In this case taking the $r_{\rm N}$=2.8 km quoted by \citet{Lamy-chapter} and calculating the nuclear magnitude at the time of their {\sl HST} observations. The uncertainty on this radius is not reported: The error bar on this point is set equal to that on INT data (0.04 mag.) for the purposes of weighting the best fit. This does lead to some uncertainty on the phase function; if the error bar is instead set equal to that from our extended NTT observations (0.007 mag.) then the corresponding increase in weight on the point gives $\beta = 0.076$ mag.~deg$^{-1}$.}, allowed calculation of a rough phase function (fig. \ref{47Pphase}). A best fit straight line gives a steep phase function of $\beta = 0.083\pm0.006$ mag.~deg$^{-1}$, implying a true $m_R(1,1,0) = 14.477\pm0.007$ and consequently $r_{\rm N}=3.63\pm0.01$ km. The fit to the three time-series measurements (taking the 5 points over 2 nights from the INT data as a `time-series') is excellent, while the snap-shot from \citeauthor{Licandro00b} also agrees with this fit if it is assumed that it was taken near a light-curve minimum. Therefore we regard this as a reasonable estimate of the phase function of 47P, despite the fact that it is unusually steep.

Colours were measured of $(V-R)  = 0.29\pm0.06$ and $(R-I)  = 0.79\pm0.07$. These are unusual when compared to typical nuclei colours, and show a large change from those presented in paper II and \citeauthor{Licandro00b}'s $(V-R) =0.4\pm0.3$. 47P was weakly active at the time of these observations, so we favour our earlier value. A weighted mean of all colour data for 47P (including that of \citeauthor{Licandro00b}) gives $(V-R) =0.42\pm0.02$, $(R-I) =0.44\pm0.03$.


\subsection{94P/Russell~4}

\begin{figure}
   \includegraphics[width=0.47\textwidth]{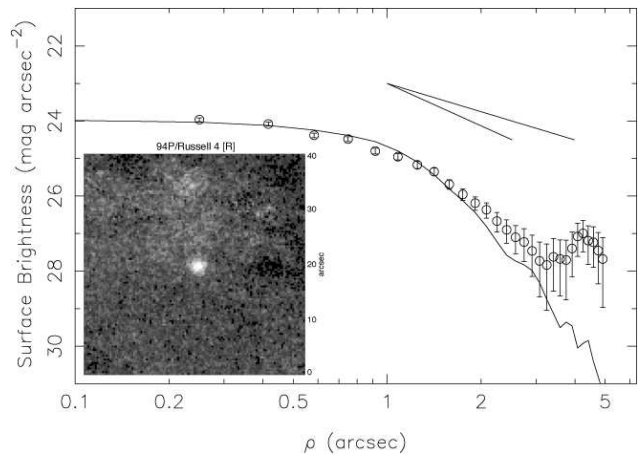}
      \caption[94P image \& SBP]{Co-added image of 24 frames taken of 94P on July 6th, 2005. Each frame had an exposure time of 75s, giving this combined frame an equivalent exposure time of $\sim$30 minutes. The profile implies that the comet could have been weakly active at the time of observation, although the rise at the end is more likely due to residual star light.
              }
         \label{94Pimage}
   \end{figure}

94P/Russell~4 was observed over 4 nights in July 2005 using the INT. A total of 83 $r'$-band frames were taken, and in addition at least one colour block on each night. The combined image and corresponding profile are inconclusive (fig. \ref{94Pimage}); the profile shows some signs of faint activity beyond 2\arcsec{} from the nucleus, although this may be due to residual flux in the combined image from particularly bright nearby stars. The calculated coma contribution within 5\arcsec{} is $m_c \ge 21.25$, or $\le 77\pm30 \%$; measured before the peak from possible stellar contamination the contribution is $m_c(3\arcsec) \ge 22.06$, or $\le 37\pm15 \%$. 

%

The mean magnitude of 94P was measured to be $m_R = 20.974\pm0.015$; as with other comets observed during this run the individual magnitudes were converted to absolute magnitudes due to the $\delta m_R \approx 0.03$ mag.~difference over the four nights, primarily due to the changing phase angle. The absolute magnitude and radius were measured to be $m_R(1,1,0) = 15.187$ and $r_{\rm N} = 2.62\pm0.02$ km. These are technically upper limits if there was any weak activity, however we believe that the comet was effectively inactive as these are in excellent agreement with preliminary results from a light-curve taken by the authors for a different program when the comet was clearly inactive at 4.7 AU in July 2007.

   \begin{figure}
   \includegraphics[origin=br,angle=-90,width=0.47\textwidth]{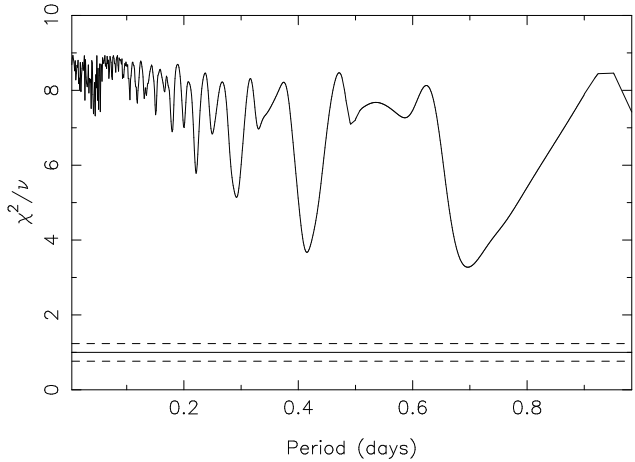}
      \caption[94P periodogram]{Periodogram for 94P. 
              }
         \label{94Ppgram}
   \end{figure}
%

   \begin{figure}
   \includegraphics[origin=br,angle=-90,width=0.47\textwidth]{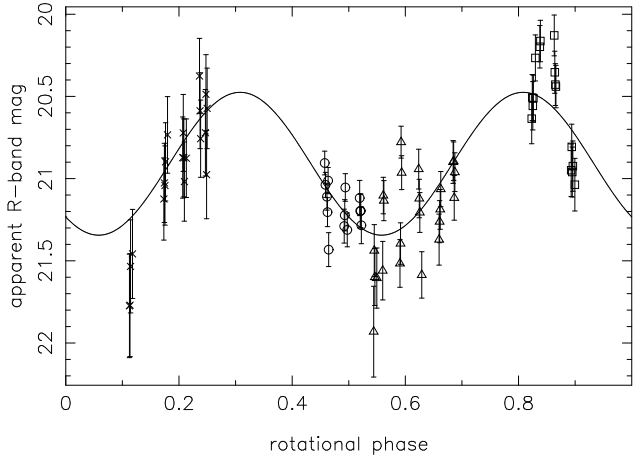}
      \caption[94P folded light-curve]{Folded light-curve for 94P, period = 33.4 hours. Data from separate nights are plotted with different symbols (circle - night 1; cross - 2; square - 3; triangle - 4): the variation in brightness is considerable even during single nights.
              }
         \label{94Pbest}
   \end{figure}
%

The motion of this comet was such that it was not possible to use common stars to produce multi-night differential light-curves, but since it was reasonably bright and exhibited a large amplitude variation the increase in size of the error bars due to calibration onto the Landolt scale was not significant. The null hypothesis of constant brightness is rejected at a $50\sigma$ level. The periodogram for 94P is shown in fig. \ref{94Ppgram}; the strongest minimum is at $P_{\rm fitted} = 16.7\pm0.4$, corresponding to a rotation period of $P_{\rm rot} = 33.4\pm0.8$ hours. Although this minimum only has $\chi^2/\nu = 193/79 = 2.45 \equiv 9\sigma$, and the data folded onto this period displays considerable scatter (fig. \ref{94Pbest}), it is clear that there is substantial variation in the brightness of the comet, again implying that the nucleus dominates the flux. 

The full range observed in the data is $\Delta m = 1.2\pm0.2$ mag., implying $a/b \ge 3$. This is quite an extreme elongation, in excess of any other measured nucleus, and exceeded by few asteroids. The fact that there are large ranges in observed brightness on individual nights and that this comet was bright show that this is not due to any calibration issue. 

An alternative explanation is that the range in $m_R$ is due to variations in albedo across the nucleus surface instead of changing cross-sectional area, which would be revealed by periodicities in the colour indices. There are some variations in the measured colours of the comet, although the individual measurements are normally distributed around the mean values of $(V-R)  = 0.62\pm0.05$ and $(R-I)  = 0.44\pm0.06$. The variations do not appear to be correlated with either of the above periods, nor can the $r'$-band data be fit to periodicities found in the $(R-I)$  sequence. Changes in surface colour do not seem to be the cause of the large amplitude variations seen in the light-curve.


\subsection{121P/Shoemaker-Holt~2}\label{121Psect}

\begin{figure}
    \includegraphics[width=0.47\textwidth]{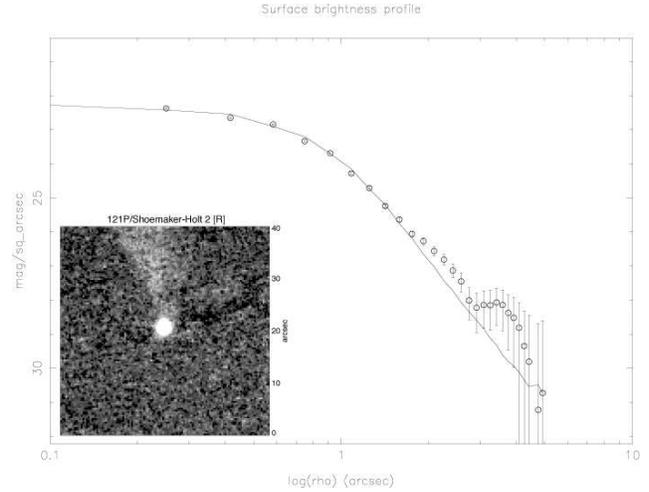}
      \caption[121P image \& SBP]{Co-added image of 121P, from the INT frames taken on 2nd March 2006. The nucleus appears star-like, and the profile implies that the comet was inactive at the time of observation, yet there is a tail visible $\sim$ West [upwards in this image] of the nucleus (highlighted in fig. \ref{121Pcolour_image}).
              }
         \label{121Pimage}
   \end{figure}

\begin{figure*}
   \begin{tabular}{c c c}
   \includegraphics[viewport=80 0 400 360,clip,width=0.32\textwidth]{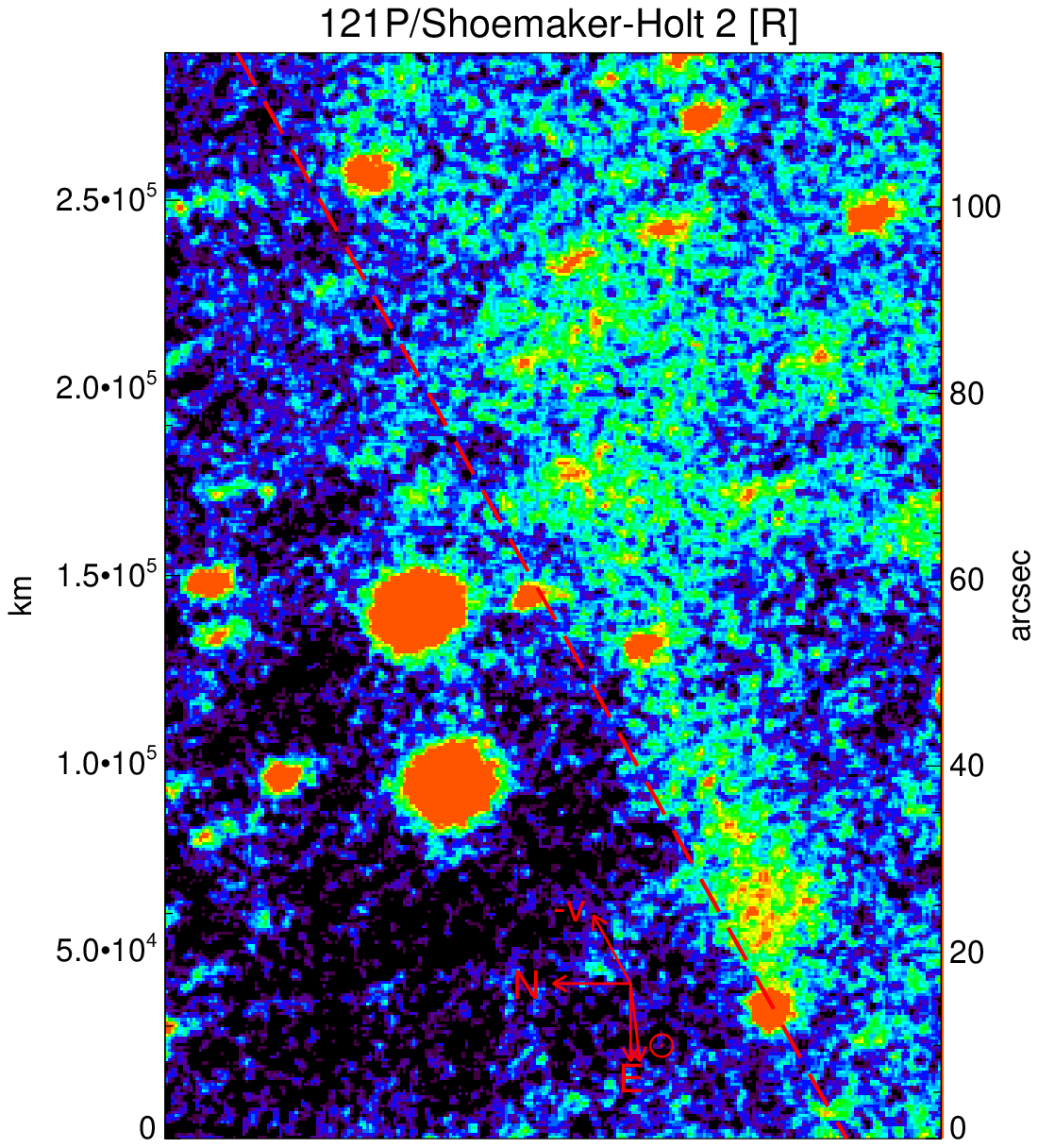}&
   \includegraphics[viewport=80 0 400 360,clip,width=0.32\textwidth]{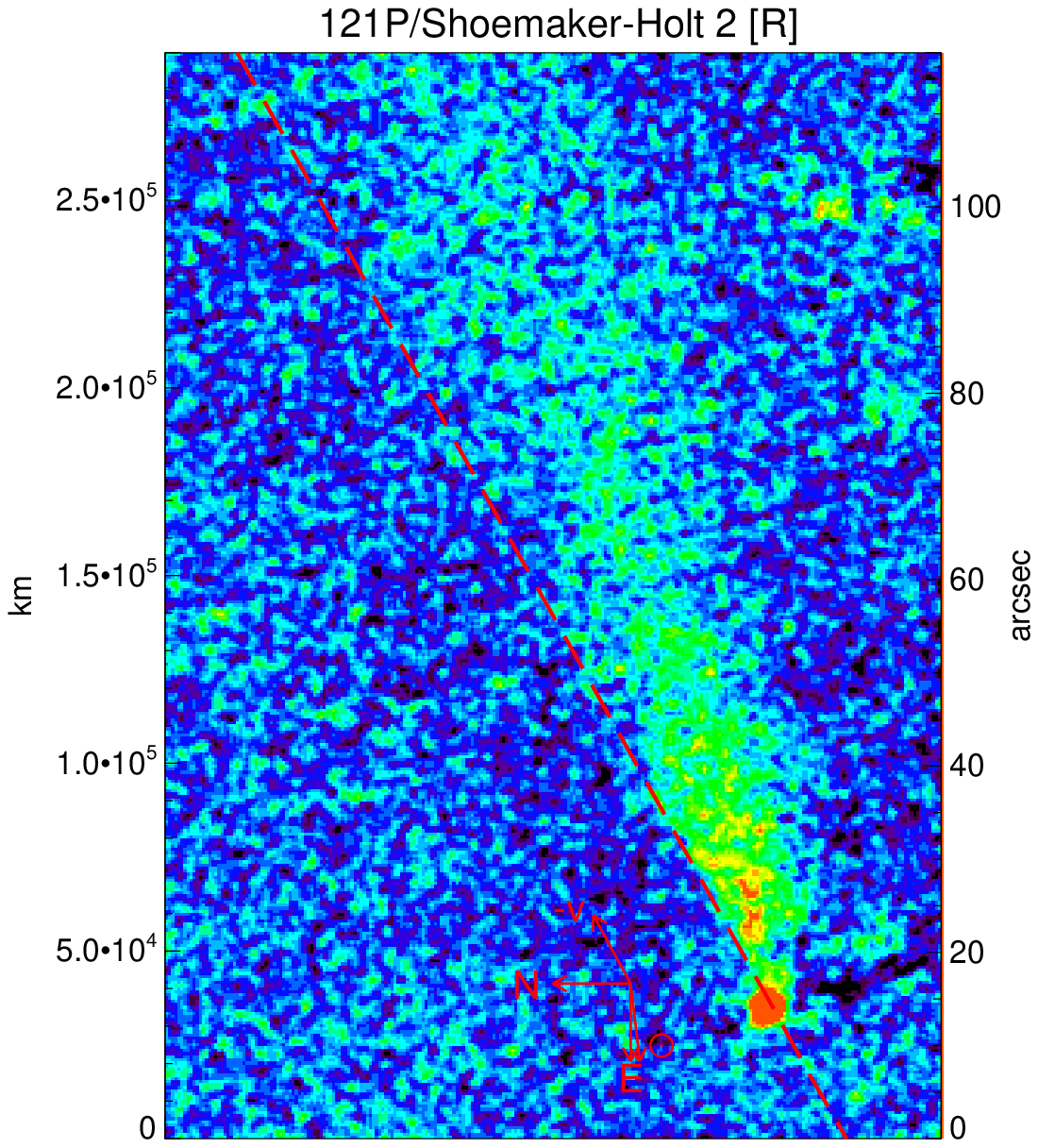}&
   \includegraphics[viewport=35 0 440 360,clip,width=0.32\textwidth]{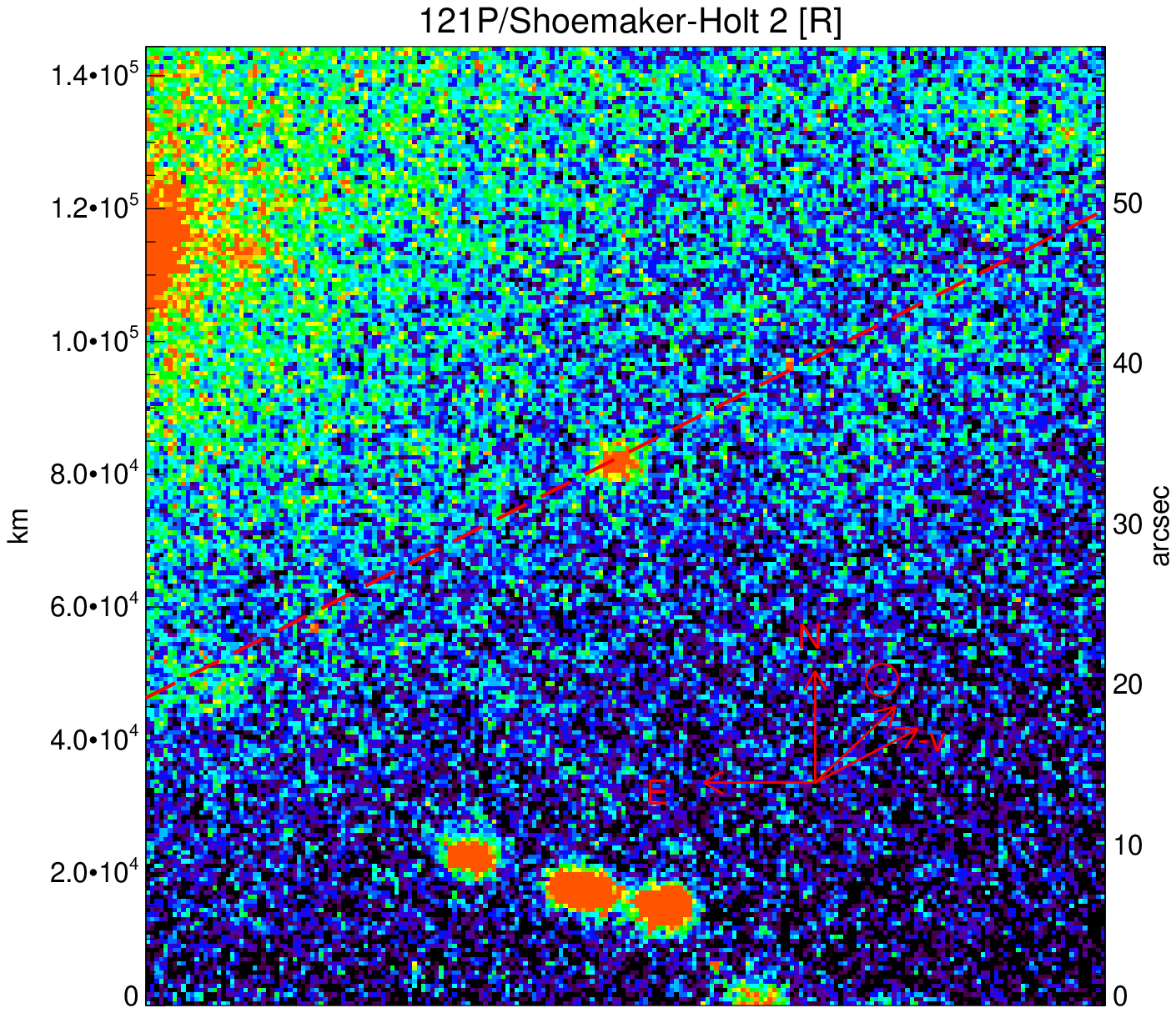}\\
   (a) & (b) & (c)
   \end{tabular}
      \caption[121P false colour images]{Wider field co-added images of 121P, using false colour to highlight the faint tail. (a) shows the 1st March 2006 INT data, during which the seeing was poor and exposure times were doubled, and there were too few images to generate a background image without the comet to subtract. This image is therefore a direct sum of the 4 $r'$-band images (equivalent exposure time $t_e = 27$ minutes), and contains stars. (b) was produced from the 18 $r'$-band frames ($t_e = 65$ minutes) taken at the INT on the 2nd of March 2006, and has had background objects removed. (c) shows the data obtained with the FTN in May 2006, and shows no tail, although it is not such a deep image (14 minutes). 
              }
         \label{121Pcolour_image}
   \end{figure*}

121P/Shoemaker-Holt~2 was predicted to have $m_R = 22.18$ during the 2006 INT run, based on a radius of $r_{\rm N} = 1.75\pm0.63$ km from a snap-shot observation in June 1999 by \citet{Lowry03}. The comet was brighter than expected, but a combined image and star-like surface brightness profile showed the comet to be inactive (fig. \ref{121Pimage}), with a calculated coma contribution within 5\arcsec{} of $m_c \ge 24.95$, or $\le 2 \pm 4\%$ of the total flux. However, inspection of the co-added frame revealed a dust feature in the plane of the comet's orbit. Originally suspected to be a dust \emph{trail} of remnant dust particles along the orbit due to the apparent inactivity of the comet, it is clear from fig. \ref{121Pcolour_image} (in which colours are used to highlight faint features, and the orbit of the comet projected onto the sky is over-plotted) that the feature is a \emph{tail}, as it is not entirely in the orbital plane and spreads around the anti-solar direction. This means that there are two contradictory measurements; the stellar nucleus profile implies that there is no coma, and there is clearly none resolvable, yet there is a tail, which implies that the comet is quite highly active. A tempting interpretation is that the observations happened to catch the comet just after it `turned off', or shortly after an outburst from an otherwise inactive comet, meaning that there is no near nucleus coma, but the material in the tail has not yet dispersed.

%

%

\begin{figure}
   \includegraphics[origin=br,angle=-90,width=0.47\textwidth]{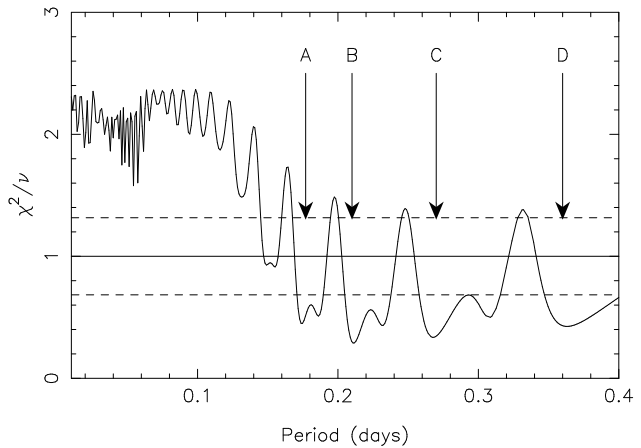}
      \caption[121P periodogram]{Reduced $\chi^2$ periodogram for 121P.
              }
         \label{121Ppgram}
   \end{figure}

\begin{figure}
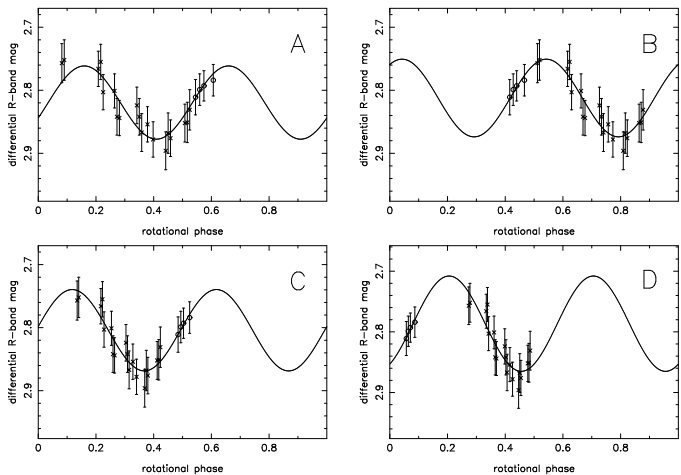

   \begin{tabular}{c c}
   \includegraphics[origin=br,angle=-90,width=0.24\textwidth]{121P-A.ps} &
   \includegraphics[origin=br,angle=-90,width=0.24\textwidth]{121P-B.ps} \\
   \includegraphics[origin=br,angle=-90,width=0.24\textwidth]{121P-C.ps} &
   \includegraphics[origin=br,angle=-90,width=0.24\textwidth]{121P-D.ps} \\
   \end{tabular}
      \caption[121P folded light-curve]{Folded differential light-curves for 121P, periods A = 8.4, B = 10.1, C = 12.8 and D = 17.4 hours. 
              }
         \label{121Pbest}
   \end{figure}

Assuming that the nucleus was inactive, the measured $m_R = 20.775\pm0.006$ implies $m_R(1,1,0) = 14.660$ and $r_{\rm N} =  3.35\pm0.01$ km. This is considerably larger than the $r_{\rm N}=1.75\pm0.63$ result found by \citet{Lowry03}, even considering the large error bar on their result. This further suggests that there is some activity present in the INT frames; the axial-ratio required to explain the difference as due to \citeauthor{Lowry03} observing an $r_{\rm N}$ = 3.35 km body at a light-curve minimum is $a/b$ = 6.3, which is unrealistic. However, the brightness of the comet does  show variations which may be periodic and could be interpreted as being due to the rotating nucleus: the periodogram and best fit folded differential light-curve are shown in fig. \ref{121Ppgram} and fig. \ref{121Pbest}. With only 4 frames taken on the first night and 19 on the second, there is ambiguity in the determined period. A number of periods give visually acceptable light-curves;  Figure~\ref{121Pbest} shows 4 that illustrate the range seen in the periodogram. Note that the deepest minima have $\chi^2/\nu < 0.68 (\equiv -1\sigma)$, implying that the error bars on the photometry are over estimated ({\it i.~e.}~the brightness measurements are more accurate than the conservative estimates on the uncertainties). The null hypothesis of constant brightness is rejected, at only $\chi^2/\nu = 47.4/21= 2.2 \equiv 3.8\sigma$, although this would be rejected at a higher confidence level if the error bars on individual points were reduced. The total variation is $\Delta m = 0.15\pm0.03$ mag., implying a minimum $a/b \ge 1.1$ and a fairly spherical shape or near pole-on orientation, although if there is near nucleus unresolved coma then the true axial-ratio will be larger (see discussion on 36P in section \ref{36Psection}). The minimum density implied by this $0.12\pm0.04$ g cm$^{-3}$.

%

%

To determine whether or not 121P was still active, and in particular to investigate the interpretation that it had just ceased outgassing in March, the comet was observed again on the 31st of May 2006, using one hour on the FTN. Based on the radius measured using the INT, the predicted magnitude at this time was $m_R =  20.69$; actually brighter than during the INT observations due to the lower phase angle, and theoretically well within the capabilities of the 2.0m FTN. 14 $R$-band frames were taken over the course of the hour, along with 3 $V$-band and suitable standard star observations for calibration. As the FTN does not have an auto-guider, exposure times were limited to \hbox{60 s} to minimise trailing of the images due to the telescope's movement. Combining all frames gave a detection of the comet, which was very close to the predicted brightness, with $m_R = 20.63\pm0.10$. The profile of this combined image indicated that the comet was inactive ($m_c(3.1\arcsec) \ge 24.2$, implying a flux contribution of $\le 4\pm7\%$), and no tail was detected to a 3$\sigma$ limiting surface brightness of $\Sigma_R \ge 25$. Unfortunately this does not rule out a tail, as the surface brightness in the centre of the tail in the INT data is beyond these detection limits at $\Sigma_R \approx 25.5$. The FTN observations were therefore unable to constrain the evolution / dissipation of the tail, although the star-like profile suggesting inactivity adds support to the idea that both sets of observations were of a bare nucleus which had recently ceased out-gassing.

If this is the case then an approximate measurement of the phase function can be made. As the measured FTN brightness was close to the predicted value, then $\beta$ must be close to the canonical 0.035; in fact a fit to these two points gives $\beta = 0.047\pm0.020$, and $m_R(1,1,0) = 14.494\pm0.006$, implying $r_{\rm N} = 3.61\pm0.01$ km. These results are still inconsistent with those of \citet{Lowry03}, and a non-standard phase law does little to improve matters. Including their result gives a phase law with $\beta = 0.036\pm0.006$, and an implied $m_R(1,1,0) = 14.65$ and $r_{\rm N}$ = 3.36, but still does not fit their magnitude well. The colours measured for 121P were $(V-R)  = 0.53\pm0.03$ and $(R-I)  = 0.44\pm0.03$, from the INT data, and $(V-R)  = 0.29\pm0.20$ in the FTN data, where the extreme faintness in the few short $V$-band exposures gives large error bars.


\subsection{P/2004~H2 (Larsen)}

\begin{figure}
    \includegraphics[width=0.47\textwidth]{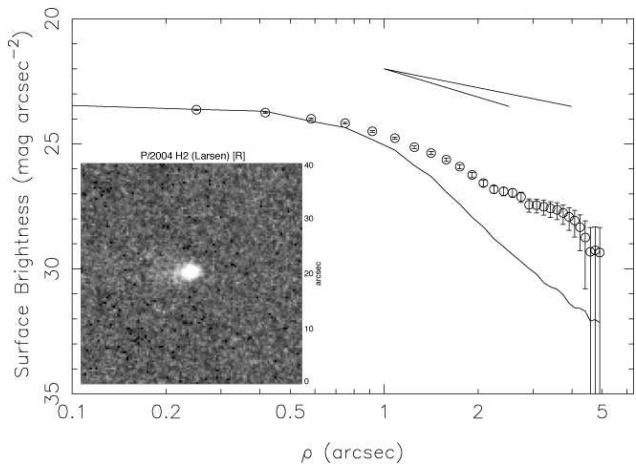}
      \caption[H2 image \& SBP]{Image showing P/2004 H2, in which a faint tail can be seen to the West (left in this image) of the nucleus. The profile also shows that the comet was weakly active, although the flux from the nucleus dominates in the photometry.}
         \label{H2image}
   \end{figure}

P/2004~H2 (Larsen) was observed in July 2005 at only its second opposition since discovery in April 2004. A total of 98 $r'$-band frames were taken; the combined image shown in fig. \ref{H2image}(a) is made up of the 16 taken on the 4th night. It can be seen in this image that the comet was weakly active, with a small tail extending to the West of the nucleus. The profile also shows the comet to be active, with $m_c = 23.85$ measured within $\rho = 5\arcsec$. The comet flux is dominated by the nucleus though, as this corresponds to only $13\pm11 \%$ of the total flux ($Af\rho = 6.0\pm0.1$ cm, within $\rho = 11,000$ km), under the assumption of steady state coma. This assumption is only an approximation in this case, as the slope of the profile has  a gradient slightly steeper than -1.5. 

P/2004~H2 was moving relatively slowly on the sky at the time of observation, allowing for both long exposure times (210 s) and the use of common stars to generate entirely differential light-curves over three night time-spans. These were searched for periodicities, as was the 7 night calibrated light-curve, adjusted to take into account the changing observational geometry. No convincing periodicity was identified.
 
As the comet was active, we find a lower limit to the absolute magnitude and an upper limit to the radius.  The light-curve does give an accurate measurement of $m_R = 21.590\pm0.007$, which reduces to an absolute magnitude of $m_R(1,1,0) \ge 15.875$ and therefore a mean effective radius of $r_{\rm N} \le 1.91\pm0.01$ km. The mean measured colours of this comet are $(V-R)  = 0.49\pm0.02$ and $(R-I)  = 0.28\pm0.04$, although the individual measurements have considerable scatter around these means, possibly due to variations in the near nucleus coma.

\begin{table*}
\begin{minipage}[]{\textwidth}

\caption[Results from time-series photometry]{Derived physical parameters and colours from time-series photometry on JFCs.}             
\label{results_lightcurve}      
\renewcommand{\footnoterule}{}  
\renewcommand{\arraystretch}{1.3}
\begin{tabular}{l c c c c c c c c c}        
\hline\hline                 
Comet & $m_R$ & $m_c$\footnote{Limiting coma magnitude measured within 5\arcsec, unless stated.} & $m_R(1,1,0)$ & $r_{\mathrm{N}}$ & $P_{\mathrm{rot}}$ & $a/b$\footnote{Lower limits as the orientation of the rotation axis is unknown.\label{fn:results_lc1}} & $D_{\mathrm{N}}$\footref{fn:results_lc1} & $(V-R)$ & $(R-I)$\\    
& & & & [km] & [hr] & & [g cm$^{-3}$] & & \\
\hline                        

36P\footnote{Results from 2005 NTT run. For each of the comets with multi-run data the individual results are presented here; the conclusions based on the combined data, including phase curves, are given in the text.\label{fn:results_lc3}} \footnote{Faint coma present. Note that these results do not include any correction for the presence of faint near-nucleus coma.\label{fn:results_lc4}}
& 21.370$\pm$0.016 & $\ge$22.4 & 15.272$\pm$0.016 & 2.52$\pm$0.02 & 3.56$\pm$0.02 & 1.4$\pm$0.1 & 1.24$\pm$0.07 & 0.46$\pm$0.05 & 0.52$\pm$0.04\\

36P\footnote{Results from 2006 INT run.\label{fn:results_lc2}} & 21.570$\pm$0.008 & $\ge$24.5 (3.3\arcsec) & 15.247$\pm$0.008 & 2.55$\pm$0.01 & $\sim$40 & 1.9$\pm$0.1 & 0.01$\pm$0.01 & 0.48$\pm$0.03 & 0.56$\pm$0.03\\

40P\footref{fn:results_lc4} & 22.08$\pm$0.02 & $\ge$22.2 & $\ge15.72$ & $\le2.05$ & n/a & n/a & n/a & 0.37$\pm$0.10 & 0.65$\pm$0.11 \\

47P\footref{fn:results_lc4}& 22.01$\pm$0.04 & $\ge$22.9 & 14.93$\pm$0.04 & 2.96$\pm$0.05 & n/a & 1.4$\pm$0.1 & n/a & 0.29$\pm$0.06 & 0.79$\pm$0.07\\

94P & 20.974$\pm$0.015 & $\ge$22.1 (3.0\arcsec) & 15.19$\pm$0.015 & 2.62$\pm$0.02 &$\sim$33 & 3.0$\pm$0.5 & 0.03$\pm$0.01 & 0.62$\pm$0.05 & 0.44$\pm$0.06 \\

121P\footref{fn:results_lc2} & 20.775$\pm$0.006 & $\ge$25.0 & 14.660$\pm$0.006 & 3.35$\pm$0.01 & 10$^{+8}_{-2}$ & 1.15$\pm$0.03 & 0.12$\pm$0.04 & 0.53$\pm$0.03 & 0.44$\pm$0.03 \\

121P\footnote{Results from 2006 FTN observations.} & 20.63$\pm$0.10 & $\ge$24.2 (3.1\arcsec) & 14.60$\pm$0.10 & 3.44$\pm$0.16 & n/a  & n/a  & n/a  & 0.29$\pm$0.20 & n/a  \\

P/2004 H2\footref{fn:results_lc4} & 21.590$\pm$0.007 & $\ge$23.9 & $\ge15.88$ & $\le1.91$ & n/a & n/a & n/a & 0.49$\pm$0.03 & 0.28$\pm$0.04 \\

\hline                                   
\end{tabular}
\end{minipage}
\end{table*}


\section{36P/Whipple}\label{36Psection}

Time-series observations were taken on 36P/Whipple on three occasions; over 1.5 nights during the March 2005 NTT run, over a further two nights a year later with the INT, and a final time with the NTT in February 2007. The reason for returning to the same comet was that, as with the others observed during the weather affected NTT run (see paper II), its rotation period was not confirmed at the first attempt. What made 36P particularly interesting was the fact that the rotation period suggested by the first data set was very short, at around 3.5 hours. Such a period implied a minimum density considerably higher than any other previously observed nucleus; equation \ref{densityeqn} gave a minimum density required for gravitational cohesion of 1.2 g cm$^{-3}$, greater than that of water ice. This preliminary result was presented at the 2005 ACM conference \citep*{Snodgrass-ACMposter}, and was noted by \citet{Toth+Lisse06}. The additional data sets do not support this result though, and show 36P's rotation period to be considerably longer. Here we first describe the results from each run before discussing the interpretation of the whole data set.

\subsection{NTT data: March 2005}

\begin{figure}
   \centering
   \includegraphics[width=0.47\textwidth]{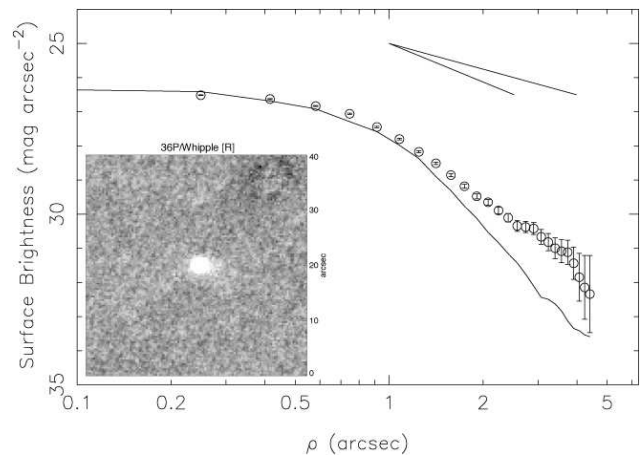}
      \caption[36P image \& SBP - NTT]{Image showing 36P, made up of 17$\times$200 second exposures taken on the 6th March 2005. There is some possible activity in the image, which appears to be confirmed by the active profile, where a clear difference between the comet and background PSF is visible.}
         \label{36Pimage}
   \end{figure}

36P was found to be bright ($m_R = 21.37 \pm 0.02$) and appeared stellar in each individual frame. A combined image (fig. \ref{36Pimage}) also appears fairly stellar, but there is a slight extension towards the East (right in fig. \ref{36Pimage}(a)) and a surface brightness profile (fig. \ref{36Pimage}(b)) shows clear signs of coma. Equation \ref{sbp_eqn} gives a coma magnitude of $m_c \ge 22.4$, which corresponds to 40 $\pm$ 23\% of the total flux being due to steady state coma. The flux within an aperture of $\rho = 5\arcsec \equiv 12000$ km gives $Af\rho = 20.2\pm0.3$, which is more properly considered an upper limit since the slope of the profile is steeper than the -1.5 limit for a steady state coma. Such activity at $R_{\rm h} > 4$ AU is surprising for a comet that has never been seen to show any large amount of activity.

\begin{figure}
   \centering
   \includegraphics[angle=-90,width=0.47\textwidth]{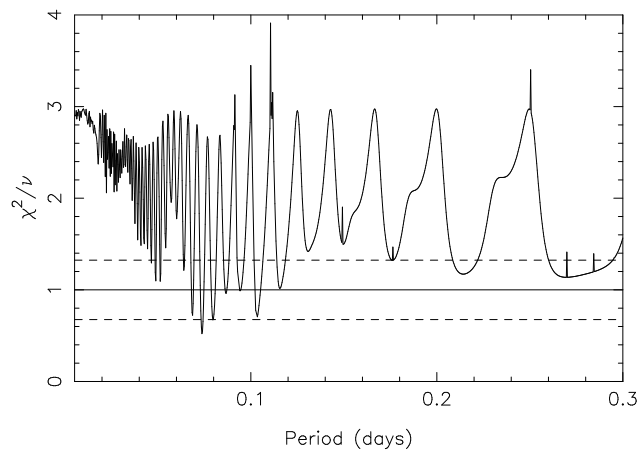}
      \caption[36P periodogram - NTT]{Periodogram for 36P, using the total observed brightness of the comet. }
         \label{36Ppgram}
   \end{figure}

\begin{figure}
   \centering
   \includegraphics[angle=-90,width=0.47\textwidth]{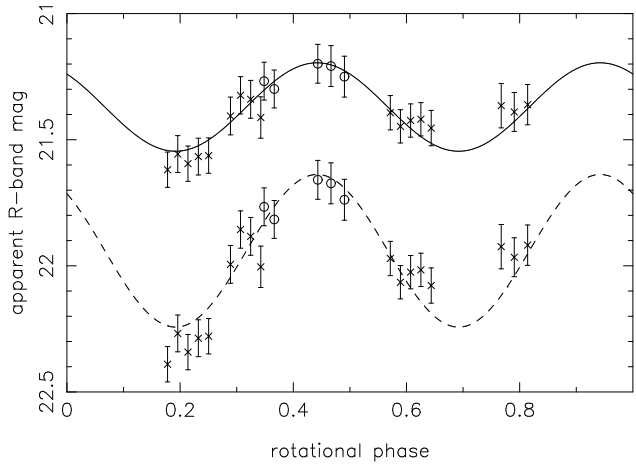}
      \caption[36P folded light-curve - coma-subtracted - NTT]{36P photometric data folded onto a 3.5 hour rotation period. The lower data set, with the dashed line drawn through it, are the data with a constant coma flux equivalent to $m_c$ = 22.4 removed from them.}
         \label{36Pncbest}
   \end{figure}

It is clear that the total flux is still dominated by the flux from the nucleus, as obvious variations can be seen in the brightness of the comet in fig. \ref{36Pncbest}, which shows the varying brightness in the original data folded onto a period of 3.5 hours. This period corresponds to twice the fitted period of 0.074 days given by the strongest minimum in the periodogram (fig. \ref{36Ppgram}); a minimum of $\chi^2/\nu$ = 0.55, which is actually below the expected value, but within $2\sigma$ of the expected $\chi^2/\nu$ = 1. The null hypothesis of constant brightness is rejected at a $6\sigma$ level, implying that the variations are real.

\begin{figure}
   \centering
   \includegraphics[angle=-90,width=0.47\textwidth]{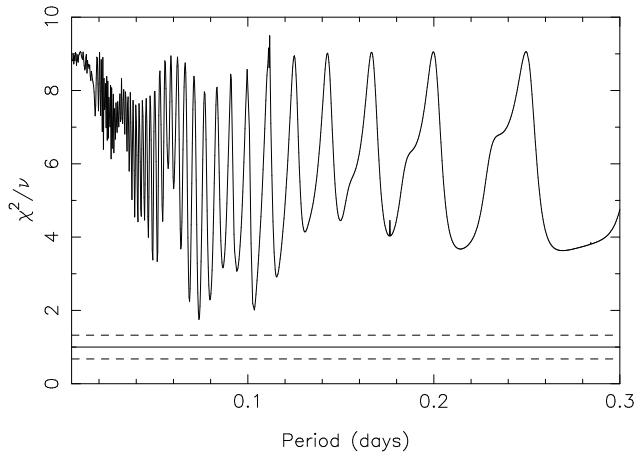}
      \caption[36P periodogram - coma-subtracted - NTT]{Periodogram for 36P, from data with a constant flux level due to the coma removed. }
         \label{36Pncpgram}
   \end{figure}

The unknown contribution to the flux from the coma causes uncertainty in both the average magnitude of the comet and the amplitude of the light-curve variation. Assuming a steady state coma, its contribution is modelled as a constant flux level, which is then subtracted from the data to leave the variation due to the nucleus. A constant coma contribution acts to raise the overall brightness, and to reduce the amplitude of the observed variation, as it will contribute relatively more when the nucleus cross section is at a minimum than at maximum. Taking the measured coma contribution from the surface brightness profile (40\% of the average total flux), and removing this from the data, gives approximate `nuclear' magnitudes. Figures \ref{36Pncpgram} and \ref{36Pncbest} show the periodogram and folded light-curve for these data. The rotation period found does not change; the presence or lack of coma at this level does not affect the measurement of this value. It can be seen that the amplitude of the light-curve is greatly increased from $\Delta m$ = 0.4 mag to $\Delta m$ = 0.65 mag, and that this increase in the size of variation relative to the size of the error bars on individual data points increases the $\chi^2/\nu$. There are no longer any minima within $\chi^2/\nu = 1 \pm \sqrt{2/\nu}$; the strongest minimum in the coma subtracted periodogram has $\chi^2/\nu$ = 1.8, corresponding to $2.3\sigma$. The mean magnitude is increased by removing a coma component; the original data has $m_R = 21.37$, removing 40\% of this flux gives $m_R = 21.94$. These values are all calculated taking the value of 40\% to be fixed; in reality the 23\% uncertainty on the flux contribution leads to large error bars on any coma subtracted photometry.

The sizes and shapes implied by the two light-curves are different. The original data give a radius for the equivalent sphere of 2.5 km, and an axial ratio of $a/b \ge 1.4$, and therefore nuclear dimensions of $a\times b = 3.3\times2.3$ km. The coma subtracted model gives a smaller, more elongated body, with $r_{\rm N} = 1.9$ km, $a/b \ge 1.8$ and thus $a\times b = 3.1\times1.6$ km. The first solution matches the value of $r_{\rm N} = 2.5 \pm 0.2$ km found by \citet{Lowry+Weissman03}, who obtained a snap-shot of 36P when it was inbound at 4.4 AU in May 2001. The second solution is also consistent with this considering the uncertainties on the coma fraction.

Figure~\ref{36Pncbest} shows that the faint coma detected does not prevent measurement of a rotation period, and that this period is independent of any corrections to the photometry applied to correct for coma. It is the determined period which is of greatest interest for this object; at 3.5 hours, these data show 36P to be the fastest rotating comet known. This implies that it has a surprisingly dense nucleus, under the assumption that it is essentially strengthless, or that this nucleus has strength not seen in others. Either of these present a challenge to the general picture of nuclei that is beginning to be built up; that they are weak and very low density bodies. The required density for a strengthless body rotating at a 3.5 hr period depends on its elongation, and therefore for 36P depends on whether or not a coma model is subtracted. Equation \ref{densityeqn} gives minimum bulk densities for 36P of 1.2 g cm$^{-3}$ for the original data (with $a/b \ge 1.4$) and 1.6 g cm$^{-3}$ for the coma subtracted version. Clearly the density is greater than water ice, for any level of coma, and over twice that required for any other nucleus.

Colours were also measured for 36P, and were found to be consistent across both nights with average values of $(V-R) = 0.46 \pm 0.05$ and $(R-I) = 0.52 \pm 0.04$. These colours are entirely typical of JFC nuclei and fall near the centre of the observed distribution, although it must be remembered that some fraction of this colour will be due to dust grains in the detected coma.

\subsection{INT data: March 2006}

\begin{figure}
   \centering
   \includegraphics[width=0.47\textwidth]{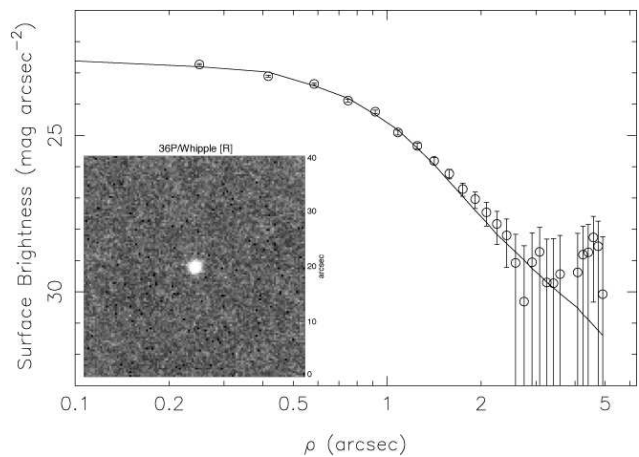}
      \caption[36P image \& SBP - INT]{Image showing 36P, made up of 55$\times$75 second exposures taken on the 2nd March 2006 with the INT, and the corresponding profile. The comet is clearly inactive.}
         \label{36PimageINT}
   \end{figure}

With this potentially interesting result in mind, 36P was the primary target of the 2006 INT run. After the first two nights were lost to poor weather, 36P was intensively monitored on the third to ensure unambiguous detection of any 3.5 hour period. A total of 63 $r'$-band images were taken over $\sim5.5$ hours, meaning that any short period would give almost complete phase coverage. As this night was non-photometric, and to search for longer periods, a considerable amount of data were also gathered on 36P on the fourth night, with 55 $r'$-band frames taken over almost 7 hours. A combination of these data and a surface brightness profile of the resultant image (fig. \ref{36PimageINT}) show that the comet was effectively a bare nucleus; a year later and 0.7 AU further from the Sun the faint activity seen in the NTT data had ceased. The formal coma limit measured within $\rho = 5\arcsec$ is $m_c \ge 23.31$, or $\le 20 \%$ of the flux, however this is probably an over-estimate due to the slight rise in the outer part of the profile due to residual sky noise (note that the profile is entirely consistent with the stellar PSF within the error bars).

   \begin{figure}
   \centering
   \includegraphics[angle=-90,width=0.47\textwidth]{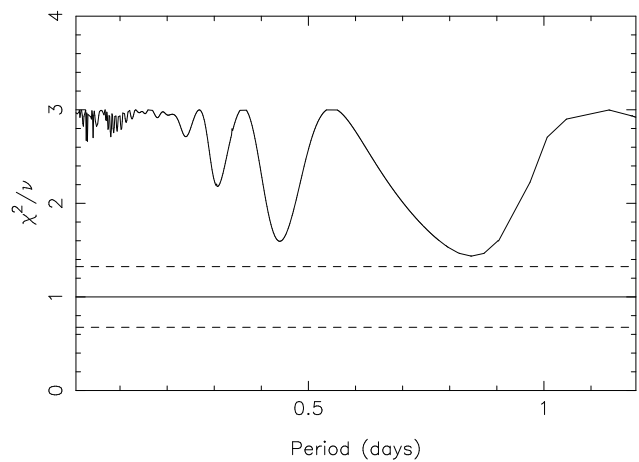}
      \caption[36P periodogram - INT]{Periodogram for 36P from INT data. }
         \label{36PpgramINT}
   \end{figure}

   \begin{figure}
   \centering
   \includegraphics[angle=-90,width=0.47\textwidth]{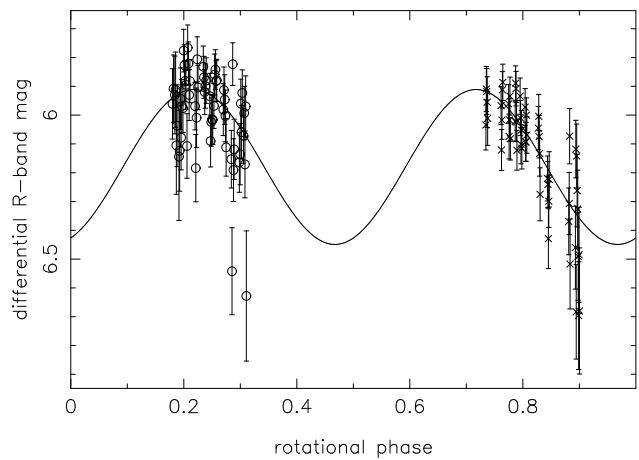}
      \caption[36P folded light-curve - INT]{36P differential photometric data folded onto an 40.6 hour rotation period.}
         \label{36PbestINT}
   \end{figure}

   \begin{figure}
   \centering
   \includegraphics[angle=-90,width=0.47\textwidth]{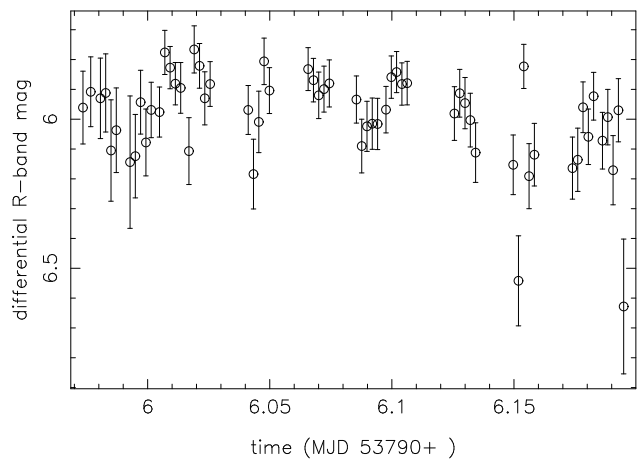}
      \caption[36P raw data - INT Night 3]{36P differential photometric data taken on the 1st March 2006.}
         \label{36PorigN1}
   \end{figure}

   \begin{figure}
   \centering
   \includegraphics[angle=-90,width=0.47\textwidth]{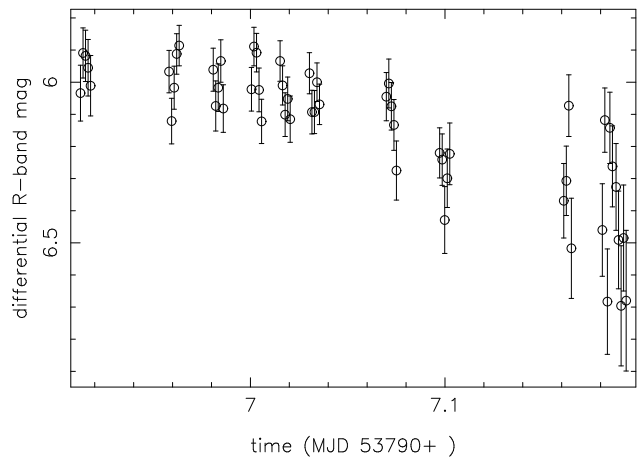}
      \caption[36P raw data - INT Night 4]{36P differential photometric data taken on the 2nd March 2006.}
         \label{36PorigN2}
   \end{figure}

Although more distant during the INT run than in the NTT run, 36P was closer to opposition, and consequently of similar apparent magnitude: $m_R = 21.570 \pm 0.008$. Assuming the standard linear phase law with $\beta = 0.035$ mag.~deg$^{-1}$ gives $m_R(1,1,0) = 15.247$ for the bare nucleus; comparison with the non-coma subtracted value from the NTT run, $m_R(1,1,0) = 15.272$, allows calculation of the fraction of flux from the nucleus during that run by:
\begin{equation}\label{nucl_percent}
10^{0.4(m_R - m_{\rm N})} = 10^{0.4(m_R(1,1,0)_{\rm NTT} - m_R(1,1,0)_{\rm INT})} = 102\%
\end{equation}
Clearly this cannot be true: the bare nucleus should be fainter than the active one, and the fraction of flux due to the nucleus should be $\sim 60\%$ if the measurement of 40\% coma contamination is correct. The discrepancy is probably due to an incorrect phase law. A discussion on the phase function for 36P is given in the next section.

The full range of the INT data is $\Delta m = 0.7\pm0.1$ mag, larger than that measured in the NTT data but similar to the range observed in the coma-corrected data. It implies $a/b \ge 1.9$. The periodogram for the INT data is shown in fig. \ref{36PpgramINT}, which shows that the fit to the fluctuations in the brightness is dominated by a long period variation. This periodicity search is based on differential magnitudes, since the large field of view of the WFC meant that the same stars could be used for comparison on both nights, meaning that no uncertainty is added due to calibration onto a standard scale. The data folded onto the best period, at $P_{\rm rot} = 40.6$ hours, is shown in fig. \ref{36PbestINT}. Such a long period could not have been detected in the shorter and sparser data set obtained with the NTT, but there remains the question of where the 3.5 hour period came from, and whether it has any basis in reality. It is clear from fig. \ref{36Ppgram} that there are many statistically acceptable periods in the NTT data; due to the few data points many periods can be found which have sufficiently low $\chi^2/\nu$. The NTT data can be folded onto a 40.6 hour period to give a visually acceptable light-curve, however this could be said of a large number of periods given the sparse data. Of more interest is whether or not any short period variation exists in the INT data. The unfolded data from each of the INT nights are shown in fig. \ref{36PorigN1} and fig. \ref{36PorigN2}, which show hints of shorter period variation. It is possible that the 3.5 hour period and these short variations are part of a non-symmetric light-curve with a longer period. The relatively noisy INT data does not allow a definite conclusion that the 40.6 hour period is correct, nor entirely rule out a short period. 

A large number of colour frames were taken within the time-series on 36P, giving good phase coverage for the fitted (long) period. Using the calibrated magnitudes of stars from the fourth (photometric) night gave the colours from differential magnitudes involving the same stars on the non-photometric night. All individual colours are consistent with means of $(V-R)  = 0.48\pm0.03$ and $(R-I)  = 0.62\pm0.02$, within the error bars on individual points. 

\subsection{NTT data: February 2007}

\begin{figure}
   \centering
   \includegraphics[width=0.47\textwidth]{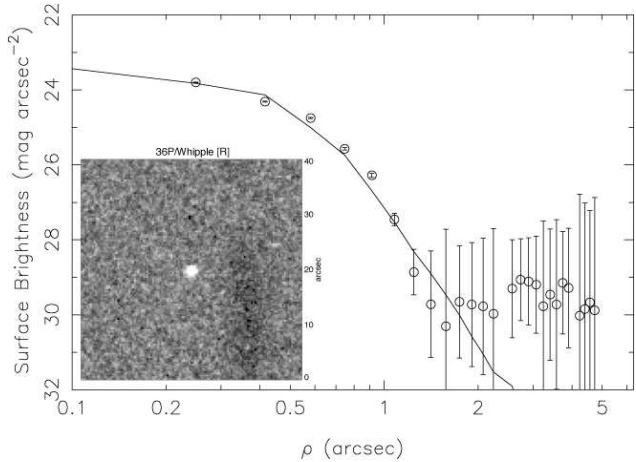}

      \caption[36P image \& SBP - DDT]{Image showing 36P, made up of 84$\times$120 second exposures taken on the 26th February 2007 with the NTT, and the corresponding profile. The comet is clearly inactive.}
         \label{36PimageDDT}
   \end{figure}

One of the main conclusions of this paper must unfortunately be that it is simply not practical to reach the required $S/N$ for accurate nucleus light-curves with a 2.5m telescope. We therefore decided to return to the 3.6m NTT to make a final search for short period variation. We were awarded 4 hours of Director's Discretionary Time (DDT) and monitored 36P continuously over 3.7 hours, and also returned to it for 15 minutes on the following night. Conditions were excellent and the comet was easily detected with $m_R = 22.774 \pm 0.013$, with the comet near aphelion at $R_{\rm h} = 5.2$ AU. The combined image and profile appear inactive (fig. \ref{36PimageDDT}), although the profile is noisy in the outer part at $\Sigma \approx 30$ mag arcsec$^{-2}$, and the formal limit is $m_c \ge 24 \pm 2$, or $\le 22 \pm 50\%$ of the flux.

   \begin{figure}
   \centering
   \includegraphics[angle=-90,width=0.47\textwidth]{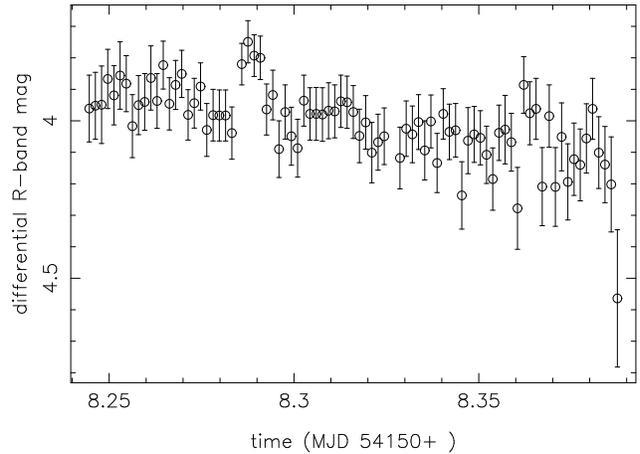}
      \caption[36P raw data - DDT Night 1]{36P differential photometric data taken on the 26th February 2007. The slight rise at $t \approx 8.29$ is due to proximity to a faint background star. Overall there is a drop of about 0.2 mag.~over the $\sim$ 4 hours.}
         \label{36PorigDDT}
   \end{figure}

There was no short period variation found in the differential light-curve (fig. \ref{36PorigDDT}). The short sequence of data on the second night was taken to search for longer period variations, but the light-curve remained fairly flat over both nights. Statistically, the null hypothesis of a constant magnitude with no variation is entirely acceptable: $\chi^2/\nu = 1.38$, within the $1\pm0.15$ range on a 1$\sigma$ variance, however there is a linear trend across the 3.7 hours of the first night, with a drop of $\Delta m \approx 0.2$ mag. We suspect that this is a short segment of a much longer period light-curve. 36P rotates with a long period; although we still cannot be sure of what this period is, it appears to be $\ge 24$ hours.

The $(V-R)$ colour of the nucleus was measured to enable calibration of the photometry including the colour term: From a single $V$-band snap-shot in the middle of the $R$-band time-series we obtain $(V-R) = 0.40\pm0.17$. A weighted average of all colour measurements gives $(V-R)  = 0.47\pm0.02$ and $(R-I) = 0.60\pm0.02$.

  \begin{figure}
   \includegraphics[origin=br,angle=-90,width=0.47\textwidth]{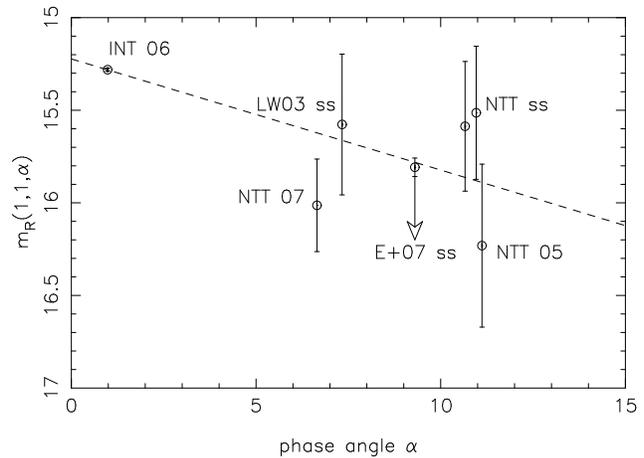}
      \caption[36P phase function]{36P mean magnitudes reduced to $r_{\rm H}=\Delta=1$ AU and plotted as a function of phase angle $\alpha$. The labels give the telescope and year of the time-series. The two labelled `NTT ss' are the 2007 snap-shots, the point `LW03 ss' is from \citet{Lowry+Weissman03} and the point `E+07 ss' is from \citet{Epifani07}, with the arrow indication that the comet was faintly active when observed by these authors and therefore this point is a lower limit. Each error bar is the combination of the photometric uncertainty on the point and the uncertainty due to only observing a section of the rotational light-curve. The error bar on the NTT 2005 point includes the uncertainty on the coma flux, which the point has been corrected for.}
         \label{36Pphase}
   \end{figure}

The three independent light-curve measurements do allow the calculation of an approximate phase function. We take the three measurements (2005 NTT at $\alpha = 11.1\degr$, 2006 INT at $\alpha = 1.0\degr$, 2007 NTT at $\alpha = 6.7\degr$), together with the snap-shot by \citet{Lowry+Weissman03} taken at $\alpha = 7.3\degr$, the deep image from \citet{Epifani07} (when the comet was weakly active at $R_{\rm h} = 3.9$ AU and $\alpha = 9.3\degr$) and two further snap-shots taken by the authors with the NTT during 2007 (at $\alpha = 10.7$ and $11.0\degr$, both at $R_{\rm h} > 5$ AU and apparently inactive). Figure \ref{36Pphase} shows this data as $m_R(1,1,\alpha)$ against $\alpha$. The error bars for each point cover the full range of uncertainty due to both the photometric uncertainty (very low for the light-curves as it decreases as $\sqrt{N}$) and also the uncertainty due to the unknown rotational phase of the observations. The maximum $\Delta m$ observed is 0.7 mag., so the snap-shots can be up to 0.35 mag.~from the mean magnitude and the DDT partial light-curve covering a 0.2 mag.~range can be up to 0.25 mag.~from the mean. The 2005 NTT data point is given with a $40\pm23\%$ coma contribution subtracted, and the error bars on this point include this considerable uncertainty. Note that we believe the INT data point to be well determined, as reflected by the small uncertainties on it, as the large amount of data over two nights covered $\Delta m = 0.7$ and thus gives a good average magnitude. 

The best fit linear phase function has $\beta = 0.060\pm0.019$ mag.~deg$^{-1}$. This is quite steep, matching the previous largest reported value for JFCs of $\beta=0.06$ for 2P/Encke \citep{FernandezY00}. We fit only a linear phase function as the data do not justify higher order fits, such as the standard IAU $(H,G)$ system \citep{Bowell89}. Also, while the INT data point is at low phase angle, we do not expect the sort of opposition surge seen in asteroids; \citet{Rabinowitz07} show that for outer Solar System bodies (TNOs and Jupiter Trojans) there is no opposition surge.

The size of the nucleus is determined from the absolute magnitudes from this phase law; extrapolating the linear phase laws back to $\alpha=0\degr$ gives $m_R(1,1,0) = 15.223\pm0.008$. Assuming a 4\% albedo this gives a radius of $2.58\pm0.01$ km. When taken with $a/b = 1.9$ this implies dimensions of $a\times b = 4.1\times 2.1 $ km.


\section{Discussion of comet properties and comparison with KBOs}\label{discussion}

Here we discuss the effect of this new data on the general trends for JFCs identified in paper II. In that paper, and in \citet{Lowry03} we suggested that the cut off in $P_{\rm rot}$--$a/b$ space, and corresponding cut off in densities, corresponded to the true density of a typical JFC nucleus, at $\sim$ 0.6 g cm$^{-3}$. If 36P was rotating in only 3.5 hours, it challenged this theory, however we now believe that it is one of the slower rotators and does not help to define the cut off. 94P also appears to be a slow rotator. The possible periods for 121P all place it well within the previously observed range, while the lack of period determinations for the other comets means that our conclusions on the rotational properties of JFC nuclei are unchanged from paper II.

\begin{figure}
   \centering
   \includegraphics[angle=-90,width=0.47\textwidth]{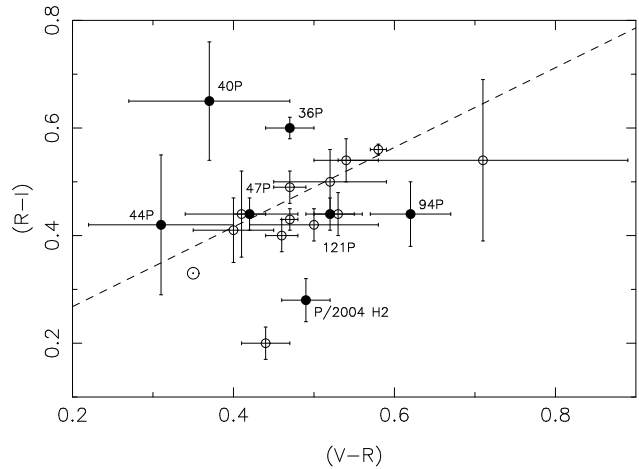}
      \caption{$(R-I)$ against $(V-R)$ for all JFCs with known colours. Filled circles -- this work; open circles -- comets with colours previously determined using the same multi-filter photometry method used here (papers I \& II and references therein). The position of the Sun on these axes is marked using the symbol $\odot$.
              }
         \label{colourplot}
   \end{figure}

The new results presented here do allow us to update the colour--colour plots presented in the earlier papers. The trend of increasing $(R-I)$ with increasing $(V-R)$ described in these papers is not as visually obvious with the inclusion of the INT data (fig. \ref{colourplot}), however this is largely due to the weakly active comets with large error bars on the colours. A weighted best fit to all the data still gives a straight line consistent with our earlier result at a 2$\sigma$ level:
\begin{equation}
(R-I)_{\rm nuc} =m (V-R)_{\rm nuc} + c \left\{ \begin{array}{l}
	m = 0.74 \pm 0.12\\
	c = 0.12 \pm 0.01
	\end{array} \right.
\end{equation}

While comets 45P/Honda-Mrkos-Pajdusakova, P/2004~H2 and 40P are not very close to this line (the first two well below the line in the lower centre of the figure, 40P above the trend in the upper left) they all showed some activity during the observations. In the case of both 40P and P/2004~H2 the coma was very weak, while 45P was observed by \citet{Lamy99} using the HST when the comet was active and at large phase angle ($\alpha \approx 90\degr$), but with a coma subtraction method. In all of these cases, the activity is not sufficient to reject the colour data as weak activity at large $R_{\rm h}$ should be dust dominated and therefore of similar optical properties to the nucleus, but also means that their difference from the rest of the data should not be over-interpreted. The mean colours of JFC nuclei are $\overline{(V-R)}_{\mathrm{nuc}} = 0.45 \pm 0.11$ ($N$ = 31) and $\overline{(R-I)}_{\mathrm{nuc}} = 0.46 \pm 0.10$ ($N$ = 19).

   \begin{figure}
   \centering
   \includegraphics[angle=-90,width=0.47\textwidth]{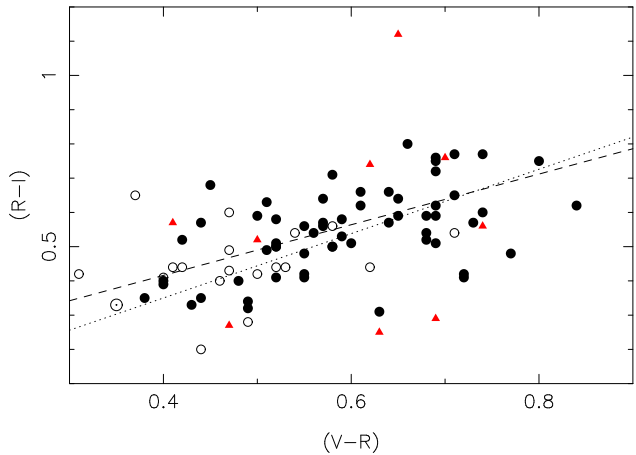}
      \caption{$(R-I)$ against $(V-R)$ for all JFCs, KBOs and Centaurs with known colours. Open circles -- JFCs; filled circles -- KBOs; triangles -- Centaurs. The position of the Sun on these axes is marked using the symbol $\odot$. The dashed line shows our best fit to the comet data, the dotted line is the fit to the KBOs. KBO and Centaur data from \citet{Jewitt+Luu01,Peixinho04}.
              }
         \label{KBOplot}
   \end{figure}

   \begin{figure}
   \includegraphics[angle=-90,width=0.47\textwidth]{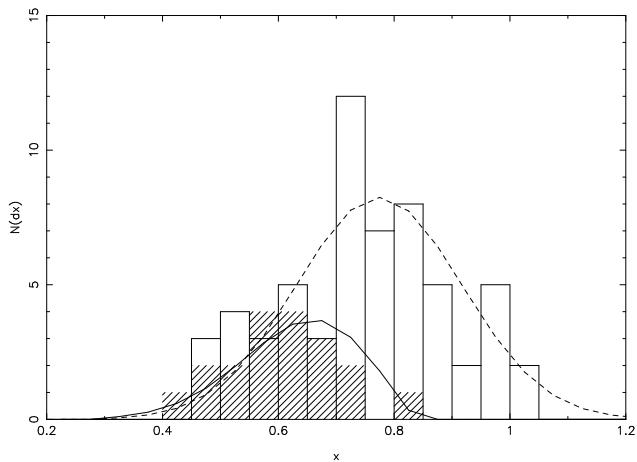}
      \caption[Colours histogram]{The $(R-I)$ against $(V-R)$ data were reduced onto distance $x$ along the best fit (see text). This plot shows the histograms of $N(x)$ against $x$ for JFCs and KBOs. The KBOs (clear) are reasonably well matched by a normal distribution (dashed line). The JFCs (hatched) appear to have a truncated distribution; the model JFC population (solid line) was created by convolving the KBO model with a `de-reddening' term that is $\propto -x$.
              }
         \label{colours_hist}
   \end{figure}

The updated fit remains consistent with the equivalent fit to KBO data (fig.~\ref{KBOplot}), while it is clear that the KBOs have redder colours. To further investigate the idea that the more neutral JFCs are `resurfaced' KBOs, it is desirable to use a one dimensional parameter to describe the relative redness of these objects. To do this the $(V-R)$ and $(R-I)$ data shown in fig.~\ref{KBOplot} were translated into distance along the best fit line $x$, which has arbitrary units linearly related to the normal colour indices, with increasing $x$ corresponding to redder colours. The weighted average best fit between the JFC and KBO best fit lines quoted above was used (with $m=0.90\pm0.05$ and $c=0.09\pm0.01$) giving a translation of
\begin{equation}
x = 0.45(V-R) + 0.67(R-I) - 0.06
\end{equation}
Histograms of the number of objects $N(x)$ per bin (d$x = 0.05$) are shown in fig.~\ref{colours_hist}. It can be seen that the JFC distribution does indeed appear to be consistent with the KBO distribution, but lacking in `ultra-red' matter. The two populations are a good match at the bluer end. The KBO colours appear to be approximately normally distributed (the fitted distribution has mean $\overline{x}_{\rm KBO} = 0.75$ and standard deviation $\sigma_{\rm KBO} = 0.14$), as do the JFCs, despite the relatively poor statistics ($\overline{x}_{\rm JFC} = 0.61$, $\sigma_{\rm JFC} = 0.10$). The model distribution plotted for the JFCs (solid line) however is not the best fit normal distribution, but one generated by convolving the fitted KBO distribution with a `de-reddening' term: $N(x)_{\rm JFC} = N(x)_{\rm KBO} \otimes f(x)$, where the fraction of objects remaining $f(x) \propto -x$. Thus the `de-reddening' depends on the colour, with the most red KBOs being most depleted, with more neutrally coloured objects remaining. A least squares fit gave $f(x) = 2.8 - 3.5x$ (again, in non-physical units) and gave a good reproduction of the observed JFC distribution. This de-reddening function means that 100\% of the KBO surfaces remain at $x=0.5$ and they are entirely depleted beyond $x=0.8$. Clearly this model is very approximate in its treatment of distributions, but it serves to demonstrate a possible transformation route between the KBO and JFC distributions. A more thorough approach would be to treat a large number of test particles, with a starting distribution like that observed in KBO colours, and then test the effect of different theoretical `de-reddening' models on the resultant JFC colours by computing the change in colour for individual comets. While an interesting problem, such calculations are beyond the scope of this work. Further colour data, especially on the JFCs, are required before a more detailed analysis of the distribution is justified.

Although a linear decrease in the number of surviving KBO surfaces with colour can be invoked to explain the observed JFC colours, a physical interpretation of this effect is required. A plausible solution is to assume that the observed variation in KBO colour is associated with the ages of their surfaces, and that space weathering reddens the surfaces in an approximately linear fashion with time. Under this paradigm the objects in very red tail of the distribution are very old, while the bluer ones have under gone some `resetting' of their surfaces ({\it e.~g.}~due to a collision) more recently. When a KBO is perturbed into the inner Solar System, the onset of cometary activity acts to resurface the nucleus. For younger KBOs, whose surfaces are not very different to their sub-surfaces as they have not been weathered for long, this has little effect, but for the very ancient and red surfaces, the difference is important. One interesting aspect of this interpretation is that it suggests that nuclei surfaces, being younger, are a closer match to their interiors, and thus may provide a better view of `primordial' material than the ancient weathered surfaces of KBOs.


\section{Summary}\label{summary}

We report here observations on JFCs from two runs with the Isaac Newton Telescope, with additional data from the Faulkes Telescope North. A large number of snap-shots were taken, and in addition time-series data were taken on 5 comets. We also describe our observations of comet 36P/Whipple over the last 3 years, at both the INT and ESO's NTT. Our results are as follows:

\begin{enumerate}
\item For the majority of the comets we have snap-shot data. Of these, five of the comets were not detected, three were visibly active, and the remainder (9 comets and one Damocloid) were apparently inactive at large distance. Most of these results were close to the detection limits of the 2.5m INT; we must conclude that good $S/N$ nucleus observations really require a 4m class telescope. The magnitudes, implied sizes, colours and activity levels for each comet are summarised in table \ref{results_snapshots}.

\item For the comets which were observed over a time-series, constraints were also placed on the rotation period and the elongation of the nucleus, from the range in magnitudes observed. Faint activity around 40P and P/2004~H2 prevented a measurement of these parameters for these comets, but for 47P, 94P and 121P and we measure minimum axial ratios of $a/b \ge 1.4, 3$ and 1.1 respectively. This implies that 94P has a very elongated nucleus. Periodogram analysis found best fit rotation periods of $\sim$ 33 and 10 hours for 94P and 121P, but both of these are very uncertain and should be regarded as estimates.

\item For 36P we present data taken at 5 epochs: 3 (partial) light-curves and 2 snap-shots. The first data set gave a preliminary result implying fast rotation for this nucleus, however larger and better sampled later data sets revealed that the rotation period is long ($\ge 24$ hours). It is still not well determined; the largest data set has a best fit period of $\sim$ 40 hours, but this is not a unique solution. This data set also shows a large amplitude variation implying an axial ratio $a/b \ge 1.9$.

\item The data on 36P were obtained at a range of phase angles $1 \le \alpha \le 11\degr$. Including the uncertainty on the snap-shot/short time series data due to the incomplete phase coverage, all the points are consistent with a linear phase function with $\beta = 0.060\pm0.019$ mag.~deg$^{-1}$. We also measure an approximate phase function for 47P, with a very steep $\beta = 0.083\pm0.006$ mag.~deg$^{-1}$.

\item $VRI$-band photometry was obtained on seven of the comets, including 36P, and the distribution of their colours was investigated. JFC nuclei and KBOs are found to have a similar trend of increasing $(R-I)$ with increasing $(V-R)$, with the JFCs being less red in general. We define a single parameter $x$ to measure the redness of the objects across the $VRI$ bands. We show that the distribution of JFC colours provides a good match to the blue end of the distribution of KBO colours, and that it is possible to recreate the JFC distribution by applying a `de-reddening' function that goes as $f(x) \propto -x$ to the KBO distribution. One interpretation of this model is that it relates to the age of the comet. 

\end{enumerate}

\section*{acknowledgements}
We wish to express our gratitude to Eduardo Gonzalez-Solares of the Cambridge Astronomical Survey Unit for providing us with pipeline processed INT data from the 2005 run. We also thank the Director General of ESO for awarding 4 hours of DDT for the final follow up observations on 36P. SCL acknowledges support from the Leverhulme trust.

\bibliography{snodgrass_etal07_INT_36P}

\end{document}